\documentclass[12pt,a4paper]{article}
\pdfoutput=1

\usepackage{amsmath,amssymb}
\usepackage[bookmarks=true]{hyperref}
\usepackage[nosort]{cite}
\usepackage[bulletsep]{collect}
\def\gfxon{\usepackage[final]{graphicx}}

\gfxon

\ifx\hypersetup\sadfkjashdfkxja\else
\hypersetup{pdftitle={Spectra of Field Fluctuations in Braneworld Models with Broken Lorentz Invariance}}
\hypersetup{pdfauthor={Peter Koroteev, Maxim Libanov}}
\hypersetup{pdfsubject={Theoretical Physics}}
\hypersetup{pdfkeywords={Large Extra Dimensions, Lorentz Invariance Violation}}
\hypersetup{plainpages=false}
\hypersetup{bookmarksnumbered=true}
\hypersetup{pdfstartview=FitH}
\hypersetup{pdfpagemode=None}
\hypersetup{colorlinks=false}
\hypersetup{citebordercolor={.5 1 .5}}
\hypersetup{urlbordercolor={.5 1 1}}
\hypersetup{linkbordercolor={1 .7 .7}}
\fi

\makeatletter
\let\old@startsection=\@startsection
\renewcommand{\@startsection}[6]{\old@startsection{#1}{#2}{#3}{#4}{#5}{#6\mathversion{bold}}}
\makeatother

\newcommand{\Mac}{\mbox{K}}

\newcommand{\dpod}[1]{\partial_{#1}}

\makeatletter \@addtoreset{equation}{section} \makeatother

\makeatletter
\let\old@makecaption=\@makecaption
\def\@makecaption{\small\old@makecaption}
\makeatother

\renewcommand{\Re}{\mathop{\mathrm{Re}}}
\renewcommand{\Im}{\mathop{\mathrm{Im}}}

\newcommand{\Integers}{\mathbb{Z}}

\ifx\genfrac\sdflkaj

\else

\fi
\newcommand{\sfrac}[2]{{\textstyle\frac{#1}{#2}}}
\newcommand{\half}{\sfrac{1}{2}}
\newcommand{\ihalf}{\sfrac{i}{2}}

\newcommand{\nln}{\nonumber\\}

\newcommand{\earel}[1]{\mathrel{}&\hspace{-2\arraycolsep}#1\hspace{-2\arraycolsep}&\mathrel{}}
\newcommand{\eq}{\earel{=}}

\makeatletter
\def\mr@ignsp#1 {\ifx\:#1\@empty\else #1\expandafter\mr@ignsp\fi}%
\newcommand{\multiref}[1]{\begingroup
\xdef\mr@no@sparg{\expandafter\mr@ignsp#1 \: }%
\def\mr@comma{}%
\@for\mr@refs:=\mr@no@sparg\do{\mr@comma\def\mr@comma{,}\ref{\mr@refs}}%
\endgroup}
\makeatother

\newcommand{\hypref}[2]{\ifx\href\asklfhas #2\else\href{#1}{#2}\fi}

\newcommand{\secref}[1]{Sec.~\multiref{#1}}

\newcommand{\appref}[1]{App.~\multiref{#1}}

\newcommand{\figref}[1]{Fig.~\ref{#1}}

\ifx\href\asklfhas\newcommand{\href}[2]{#2}\fi
\newcommand{\arxivno}[1]{\href{http://arxiv.org/abs/#1}{#1}}

\begin{document}

\begin{flushright}\footnotesize
\texttt{ArXiv:\arxivno{0901.4347}}\\
\texttt{ITEP-TH-31/08}\\
\texttt{ULB-TH/08-37}\\
\vspace{0.5cm}
\end{flushright}
\vspace{0.3cm}

\renewcommand{\thefootnote}{\arabic{footnote}}
\setcounter{footnote}{0}
\begin{center}%
{\Large\textbf{\mathversion{bold}
Spectra of Field Fluctuations in Braneworld Models\\
with Broken Bulk Lorentz Invariance}%
\par}

\vspace{1cm}%

\textsc{Peter Koroteev$^{\ddag,\ast,\star}$\, and\, Maxim Libanov $^{\ddag}$}

\vspace{5mm}

\textit{$^{\ddag}$Institute for Nuclear Research of the Russian Academy of
Sciences\\
60th October Anniversary Prospect 7a, Moscow 117312, Russia}

\vspace{5mm}%

\textit{$^{\ast}$Institute for Theoretical and Experimental Physics\\%
 B. Cheremushkinskaya 25, Moscow 117218, Russia}

\vspace{5mm}%

\textit{$^{\star}$University of Minnesota, School of Physics and Astronomy\\%
116 Church Street S.E. Minneapolis, MN 55455, USA}

\vspace{5mm}

\thispagestyle{empty}

\texttt{koroteev@physics.umn.edu, ml@ms2.inr.ac.ru}

\par\vspace{1cm}

\vfill

\textbf{Abstract}\vspace{5mm}

\begin{minipage}{12.7cm}
We investigate five-dimensional braneworld setups with broken Lorentz
invariance continuing the developments of \cite{Koroteev:2007yp}, where
a family of static self-tuning brane\-world solutions was found. We show
that several known brane\-world models can be embedded into this family.
Then we give a qualitative analysis of spectra of field fluctuations in backgrounds with broken
 Lorentz invariance. We also elaborate on one particular model and study spectra of
scalar and spinor fields in it. It turns out that the spectra we have found
possess very peculiar and unexpected properties.
\end{minipage}

\vspace*{\fill}

\end{center}

\newpage

\section{Introduction and Overview}\label{sec:Intro}

In modern cosmology and particle physics we are faced with several
problems like the cosmological constant problem and the hierarchy
problem. Some time ago the novel braneworld
programme was proposed \cite{Rubakov:1983bb, ArkaniHamed:1998rs,
Randall:1999ee, Randall:1999vf} which made the above problems more
tractable \cite{Rubakov:1983bz, Antoniadis:1998ig, Sundrum:1998ns, Bajc:1999mh,
ArkaniHamed:2000eg, Csaki:2000wz, Dvali:2002pe, Gunther:2003zn,
Burgess:2005wu, Burgess:2007ui, Carroll:2003db} (see also
\cite{Witten:2000zk} for string theoretical analysis). In this treatment
our Universe is represented as a four-dimensional submanifold (a brane)
embedded into a manifold of higher dimension (a bulk). The presence of
extra dimensions changes cosmological evolution
\cite{Binetruy:1999ut, Binetruy:1999hy} and could be made manifest at
forthcoming high energy experiments \cite{Rizzo:1998fm, Gupta:2004vr,
Vacavant:2003jb, Benslama:2004ta, Pankov:2004rt, Dvergsnes:2004tc,
Ghezzi:2005wx, Brochu:2005tr, Dannheim:2006wd, Agashe:2006hk,
Shmatov:2007mg, Berry:2007fk, Melbeus:2008hk} and in the other
measurements \cite{Agashe:2001ra, Adelberger:2002ic, Long:2003ta,
Frere:2004yu}. Among an enormous number of the large extra dimensional
models the Randall--Sundrum models (RS1 \cite{Randall:1999ee} and the RS2
\cite{Randall:1999vf}) appeared to be very successful, as they
allow one to look at the above two problems from a new viewpoint
without spoiling four-dimensional physics at observable distances and
energies. The RS models have initiated the extended and prospective
programme of large extra dimensional physics; it is not easy to present an
exhaustive list of publications devoted to physics in extra dimensional
setups, one can use various reviews and notes for finding necessary
literature \cite{Rubakov:2001kp, Gabadadze:2003ii, Csaki:2004ay,
Brax:2004xh}.

In the Randall--Sundrum models the five dimensional anti de-Sitter space
is used as a bulk filled by the negative cosmological constant and one (RS2)
or two (RS1) branes represented by 3+1 dimensional hypersurfaces embedded
into the bulk. One can also consider various generalizations of this
configuration and study their phenomenological manifestations. One of the
possible extensions of the RS models consists of an inclusion of an energy
density (matter) in consideration which is somehow distributed in the bulk
and/or on the brane. The presence of matter may lead in general to
breaking of Lorentz invariance. Different aspects of Lorentz invariance
violation in models with large extra dimensions were considered in
\cite{Rizzo:2005um, Baukh:2007xy, Gorbunov:2005dd, Gorbunov:2008dj, Obousy:2008ca}. In
these models Lorentz invariance was broken spontaneously on the brane due
to the presence of a symmetry-breaking potential. However, one can also
consider models with broken Lorentz invariance in the bulk and study
deviations from the four-dimensional physics on the brane which \textit{a
priori} preserves Lorentz invariance (see, e.g., \cite{Dubovsky:2001fj,
Libanov:2005yf, Libanov:2005nv, Ahmadi:2006cr, Koroteev:2007yp}). It is
worth noting however that, as it was proved in~\cite{Koroteev:2007yp}, the matter
located in the bulk which produces a Lorentz violated static smooth
metric, inevitably violates the null energy conditions.

In the present work we investigate field fluctuations in the backgrounds
found in \cite{Koroteev:2007yp}, where the bulk metric explicitly breaks
Lorentz invariance and has the following form
\begin{equation}\label{eq:GeneralansatzIntro}
ds^2=\mathrm{e}^{-2a(z)}dt^2-\mathrm{e}^{-2b(z)}d\textbf{x}^2-dz^2\,.
\end{equation}
Here $(t,\textbf{x})$ are four-dimensional coordinates parallel to the
brane and $z$ is the coordinate along the extra dimension. In
\cite{Koroteev:2007yp} a two-parametric family of solutions of the bulk
Einstein equations has been found in the presence of an ideal anisotropic
relativistic fluid in the bulk.\footnote{In accordance with the theorem
proved in~\cite{Koroteev:2007yp}, this fluid violates null energy
conditions.} The functions $a$ and $b$ from (\ref{eq:GeneralansatzIntro})
are linear functions of $z$
\begin{equation}
\label{eq:abxizeta}
a(z)=\xi k|z|,\quad b(z)=\zeta k|z|,\quad k>0\,,
\end{equation}
with the coefficients $\xi$ and $\zeta$ related to the anisotropy
parameters of the ideal fluid. In this paper we introduce a space
parametrized by $\xi$ and $\zeta$ which describes different metric tensors
and study how the fluctuation spectra of fields behave at different values
of $\xi$ and $\zeta$.

First, we perform more generic analysis of the spectra employing the metric
(\ref{eq:GeneralansatzIntro}) without specifying the functions $a$ and
$b$. One of our main results is that the spectrum pattern is
governed by an asymptotic behavior of the functions $a$ and $b$. If
$a(z)/b(z)\to 0$ as $z\to \infty$ then the spectrum is \textit{discrete}
and all excitation modes are \textit{localized} in some vicinity of the
brane located at $z=0$. The reason of the localization is that in this
case the potential responsible for the localization grows away from the
brane. In the opposite case, if $a(z)/b(z)$ does not vanish at infinity,
then at least part of the spectrum becomes \textit{continuous} and
excitation modes with energy high enough \textit{delocalize} from the
brane. In this case the potential tends to some constant value (or
vanishes) far away from the brane. We further embed some known models: flat
Minkowski space, the RS2 model \cite{Randall:1999vf}, Dubovsky model
\cite{Dubovsky:2001fj}, into our parametric solution.

In addition, we elaborate on a new model with the metric
(corresponding to $\xi=0,\,\zeta=1$ in (\ref{eq:abxizeta}))
\begin{equation}
\label{eq:MetricOur}
ds^2=dt^2-\mathrm{e}^{-2k|z|}d\textbf{x}^2-dz^2\,,
\end{equation}
where $g_{00}$ metric coefficient is unwarped, and  $g_{ij}$
coefficients are warped in the same way as in the RS
model\footnote{Five-dimensional coordinates $x^A$ are labeled by capital
Latin indices 0,1,2,3,5; three-dimensional coordinates $x^{i}$ are labeled
by lower Latin indices 1,2,3.}. The model with metric (\ref{eq:MetricOur})
is shown to have dramatically different properties than the models
\cite{Randall:1999vf, Dubovsky:2001fj} as long as localization of
perturbations and speed of particle propagation are considered. In the
former field perturbations are localized near the brane and their spectrum
is discrete, while for the latter there may exist only several
(quasi-)localized modes and the spectrum is continuous. We study spectra
of scalar and spinor excitations in the background (\ref{eq:MetricOur}).
In both cases we find that there is a zero (or gapless) mode --- a
state with vanishing energy  as three-dimensional momentum tends to zero.
From the four-dimensional point of view a zero mode corresponds to a
massless state and this is the only state which presents in the low energy
effective theory. It is remarkable that in spite of the metric
(\ref{eq:MetricOur}) is Lorentz invariant on the brane ($z=0$) the zero
mode dispersion relation explicitly breaks Lorentz invariance at low
momenta because of the ``tail'' of the zero mode wave function ``feels''
the breaking of Lorentz invariance in the bulk.

The paper is organized as follows. In \secref{sec:LIVSummary} we make some
qualitative remarks about field fluctuation spectra in different
backgrounds with broken bulk Lorentz invariance. In \secref{sec:Scalars}
and \secref{sec:Fermions} we investigate spectra of scalar and spinor
fields respectively in these backgrounds. Our conclusions and final
comments are given in \secref{sec:Conclusions}. Detailed derivation of the
fermionic spectrum is presented in \appref{sec:Derivation}.

\section{Localization and Delocalization of Fluctuations}
\label{sec:LIVSummary}

In this section we show that the asymptotic behavior of the functions
$a$ and $b$ in (\ref{eq:GeneralansatzIntro}) is responsible for the pattern of perturbation spectra of fields in the backgrounds in question. In the next sections we investigate how
different perturbations behave in a particular example. From
qualitative description one can understand something about those
perturbations in a generic fashion.

Considering some bulk matter field one can think of (quantum) particles
propagating in the bulk. Let us assume that our model admits
semi-classical approximation. If so, then particles do not
considerably deviate from their classical trajectories and become
``localized'' in domains with higher density of these trajectories. Thus
investigation of a geodesic configuration helps us in understanding of
localization properties of field fluctuations. For simplicity we
consider a massless free field, so the corresponding particles propagate
along null geodesics.  A geodesic is a solution to the following equation
\begin{equation}
\label{eq:GeodesicEquation}
\ddot{x}^A+\Gamma^A_{BC}\dot{x}^B\dot{x}^C=0\,,
\end{equation}
where $\Gamma^A_{BC}$ are Christoffel symbols for the metric
(\ref{eq:GeneralansatzIntro}) and the dots above $x^A$ stand for
derivatives with respect to the affine parameter.
Equations for the $0,1,2,3$ components can be integrated straightforwardly
\begin{equation}
\dot{x}^0 = \alpha\,\mathrm{e}^{2a},\quad \dot{x}^i =
\beta^{i}\,\mathrm{e}^{2b},\quad i=1,2,3\,,
\end{equation}
where $\alpha$ and $\beta^{i}$ are the integration constants. Using this
fact and light cone equation for the massless particle $g_{AB}\dot
x^{A}\dot x^{B}=0$ one finds the particle velocity along fifth direction
\begin{equation}
\dot z^{2}=\alpha ^{2}\mathrm{e}^{2a}-\left(\beta^{i}\right)^2\mathrm{e}^{2b}\,.
\label{Eq/Pg6/1:fluctpdf_0801}
\end{equation}
It follows from this equation that if the three-dimensional initial
velocity vanishes, $\beta _{i}=0$, then the particle motion is always
non-finite. From the quantum mechanical point of view it means that the
particle with zero three-dimensional momentum cannot be localized near the
brane. The particle motion is also non-finite in the Lorentz invariant
case, $a(z)=b(z)$. It does not, however, contradict the fact of presence
of a zero mode in the RS2 setup --- the semi-classical description is
certainly not applicable for a zero mode.

Let us consider the case $a(z)\to +\infty $ and $b(z)\to +\infty $ as
$z\to \infty $. This is the most general case corresponding to the finite
volume extra dimension. In this case if $a(z)/b(z)\to 0$ as $z\to \infty $
then the particle motion is finite ($\beta _{i}\neq 0$) which corresponds
to localization of the quantum particle in some domain near the brane. The
opposite case $a(z)/b(z)\to \infty $ or $a(z)/b(z)\to \mathrm{const}$ as
$z\to \infty $  corresponds to the delocalization regime and there exists
a probability of a detection of the particle far away from the brane.

We would like to mention again that the considerations used in this
section are merely qualitative. An advantage of this description is that
its conclusions are universal and can be applied to any field theory in the bulk.

\section{Scalar Fluctuations}\label{sec:Scalars}

After qualitative analysis in the previous section let us proceed to more
specific treatment and investigate fluctuations of a scalar field in the
bulk. For simplicity in this section we consider massless scalar field.
First, we make a qualitative description of how the perturbations behave
for generic form of the metric (\ref{eq:GeneralansatzIntro}) and discuss
some special points in the space parameterized by $\xi$ and $\zeta$. In the end we elaborate on the
spectrum of the scalar field in the background (\ref{eq:MetricOur}).

Let us consider the massless scalar field with the five-dimensional
action
\[
S = \int dt
d\textbf{x}\int\limits_{-\infty}^{+\infty}dz\,\sqrt{g}\,
g^{AB}\partial_A\phi \partial_B\phi\,.
\]
We assume that presence of the field $\phi$ in the bulk does not affect
the background metric. The equation of motion reads
\[
\left[-\dpod{z}^2+(a'+3b')\dpod{z}+\mathrm{e}^{2a(z)}\dpod{t}^2-
\mathrm{e}^{2b(z)}\partial^2_i\right]\phi=0\,,
\]
and, after four-dimensional Fourier transform
\[
\phi(t,\textbf{x},z)=\frac{1}{(2\pi )^2}\int dE\,d\textbf{p}
\,\mathrm{e}^{iEt-i\textbf{p}\textbf{x}} \phi(E,\textbf{p},z) \,,
\]
it takes the following form
\[
\left[\dpod{z}^2-(a'+3b')\dpod{z}+E^2\mathrm{e}^{2a(z)}-
p^2\mathrm{e}^{2b(z)}\right]\phi=0\,,
\]
where we denoted $p=\sqrt{\textbf{p}^2}$. For convenience we introduce
another function
\[
\chi(z)=\exp\left(-\half a-\sfrac{3}{2}b\right)\phi(z)\,,
\]
and the above equation can be rewritten as follows
\begin{equation}
\label{eq:KleinGordonEquationz}
\left[\dpod{z}^2+E^2\mathrm{e}^{2a(z)}-p^2\mathrm{e}^{2b(z)}+
\frac{a''+3b''}{2}-\frac{(a'+3b')^2}{4}\right]\chi=0\,.
\end{equation}
We also introduce the new coordinate $y$ by the relations
\[
\frac{\partial}{\partial z}y(z)=\mathrm{e}^{a(z)},\quad
y(z)=\int\limits_0^z \mathrm{e}^{a(\rho)}d\rho\,.
\]
In terms of this new coordinate, the equation
(\ref{eq:KleinGordonEquationz}) takes the following form
 \begin{equation}
\label{eq:SchroedingerEquation}
\frac{\partial ^{2}\chi }{\partial y^{2}}+(E^{2}-V)\chi=0\,,
\end{equation}
where
\begin{equation}
\label{eq:EnergyPotential}
V = -\frac{1}{4} \left(\frac{\partial a}{\partial y}   \right)^{2} +
\frac{9}{4}\left(\frac{\partial b}{\partial y}   \right)^{2} +p^2
\mathrm{e}^{2(b-a)}-\frac{1}{2}\left(\frac{\partial^{2} a}{\partial
y^{2}}+3\frac{\partial^{2} b}{\partial y^{2}}\right)\,.
\end{equation}

Let us consider the potential $V$ in more
detail. In what follows we assume $\Integers_2$ bulk reflection symmetry
$z\to -z$ with respect to the brane. It means in particular that the
functions $a$ and $b$ in (\ref{eq:GeneralansatzIntro}) depend on the
absolute value of $z$ (and, hence, on $y$) only. The potential $V$ is
bilinear in the first and linear in the second  derivatives, hence in
general it has a delta-function term emerging from the last term in
(\ref{eq:EnergyPotential}). The sign of the last term can be either
positive or negative depending on the behavior of the functions $a$ and
$b$ in the vicinity of the origin. If the sign is positive then we have a
delta-well in the potential and a delta-peak otherwise. If Lorentz
invariance is not broken ($a(y)=b(y)$) then the momentum-dependent term is
merely a constant and does not affect the shape of the potential $V$.
However, if this is not the case, depending on the difference $b-a$ in the
exponent the shape of the potential can be drastically modified. If $b>a$
as $y$ increases into the bulk, then the momentum-dependent term increases
as well and soon will prevail the other terms.\footnote{We do not pay much
attention here on rigorous analysis of conditions how the functions $a$
and $b$ affect the potential, leaving it to the next subsection where an
explicit example is elaborated.} The opposite configuration with $b<a$
makes (far enough from the brane) the momentum-dependent term negligible
compared to the others. Thus  we see that the
generic potential (\ref{eq:EnergyPotential}) can be classified by the
following properties
\begin{itemize}
\item Behavior at infinity which is controlled by the Lorentz invariance
violation (the sign of $b-a$). The potential $V$  increases/decreases as
$y\to\infty$.
\item Sign of the delta-function term. The potential may have either
delta-well or delta-peak at the origin depending on this sign.
\end{itemize}
This rough classification enables us to describe the behavior of
fluctuations in each case.
\begin{itemize}
\item If the momentum-dependent term increases as $y\to\infty$ the
potential has a box-type shape and the excitation spectrum is discrete.
The potential may have local minima and maxima, but qualitative pattern of
the spectrum is determined by its behavior at infinity. On the contrary,
if the potential decays at infinity, then we have continuous spectrum of
plane waves propagating along $y$-direction. Some combination of these two
scenarios is possible if $V\to V_{\infty}=\mathrm{const}$ as $z\to\infty$.
Then those modes with $E^{2} < V_{\infty}$ belong to discrete spectrum and
modes with $E^{2} > V_{\infty}$ contribute to continuous spectrum.

\item The sign of the delta-function term affects zero mode existence. In
a delta-well there might be a zero-mode\footnote{There might be also no
zero mode, but one cannot understand whether it exists or not without
detailed analysis.} and its existence is very unlikely in a configuration
with a delta-peak.
\end{itemize}

Unfortunately rigorous
analysis of (\ref{eq:SchroedingerEquation}) with arbitrary functions
$a$ and $b$ is troublesome as the corresponding equations cannot be solved
analytically. Nevertheless, the qualitative behavior of
perturbations is not changed if we consider metrics, different in
general, but having similar form in the vicinity of the origin and at
infinity. In \secref{sec:ModelA} we choose the functions $a$ and $b$ to be
of the form (\ref{eq:abxizeta}) which enables us to obtain exact solutions
for certain configurations.

\subsection{Parameter Space}

In \cite{Koroteev:2007yp} the static solution for the ansatz
(\ref{eq:GeneralansatzIntro}) has been found in the presence of an
anisotropic perfect fluid in the bulk. The metric has the following form
\begin{equation}
\label{eq:metricxizeta}
ds^2=\mathrm{e}^{-2\xi k|z|}dt^2 -
\mathrm{e}^{-2\zeta k|z|}d\textbf{x}^2- dz^2\,.
\end{equation}
Obviously, the chosen parametrization with $\xi,\zeta$ and $k$ is
overcomplete,\footnote{For instance, one could impose the constraint
$\xi^2+\zeta^2=1$ and let $k$ play a role of curvature scale.} one of the
parameters can be scaled away. Nevertheless, in what follows we keep all
three parameters unconstrained in order to simplify further analysis.

\begin{figure}[!ht]
\unitlength=1mm
\begin{picture}(110,110)(-30,-5)
\includegraphics[height = 10cm, width=11cm]{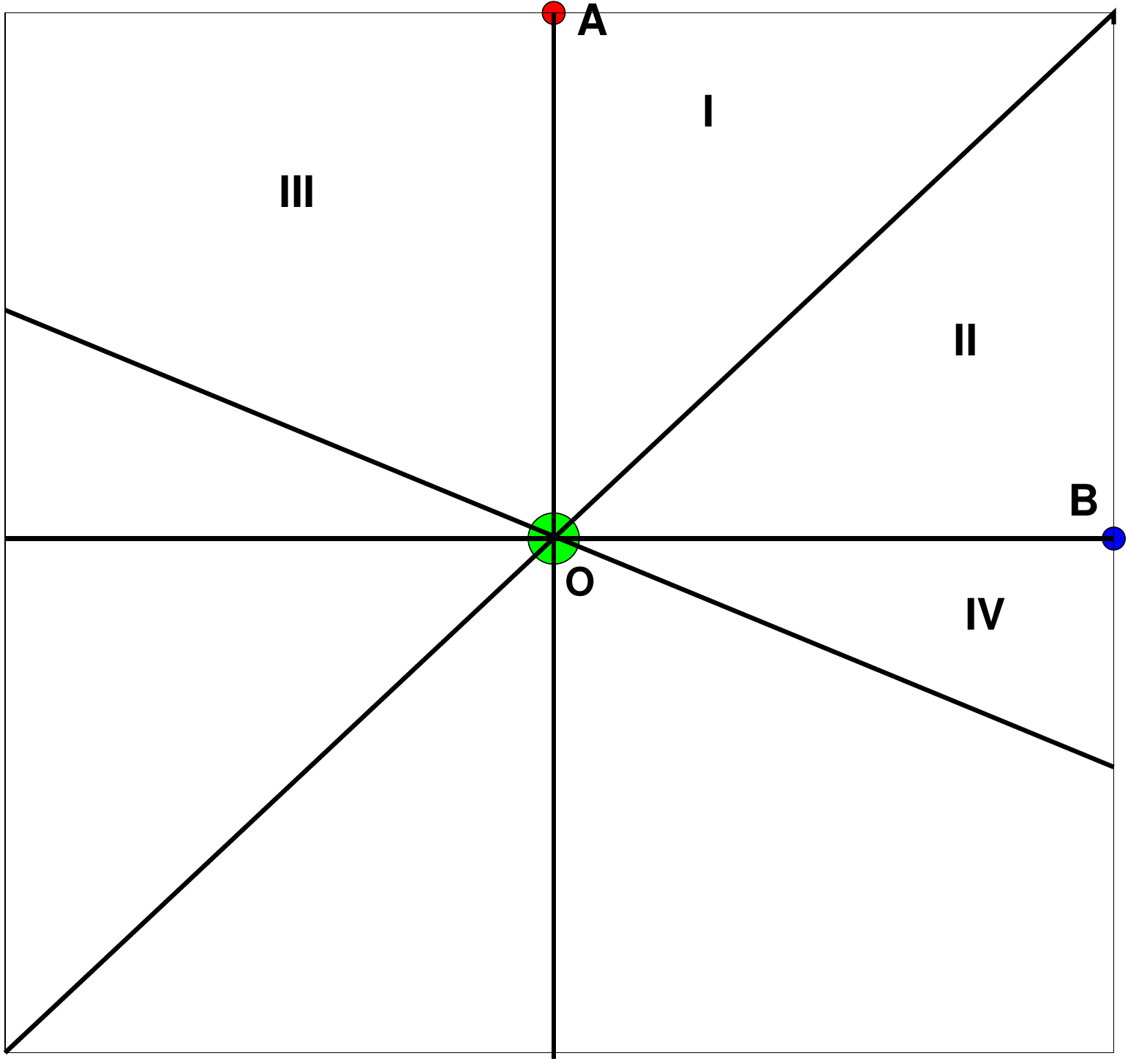}
\put(-5,43){$\xi$}
\put(-59,93){$\zeta$}
\end{picture}
\caption{The parameter space. We discuss our theory in the following
subsets of this square: Minkowski (the origin O), $AdS_5$ (the diagonal),
type I (upper triangle) and type II (lower triangle).}
\label{fig:xizeta}
\end{figure}

One can depict the $(\xi,\zeta)$ parameter space as shown in \figref{fig:xizeta}.
One has flat Minkowski metric at the origin O, $AdS_5$ metric on the main
diagonal, the models with $\xi<\zeta$ (type II models) in the lower triangle
and models with $\xi>\zeta$ (type I models) in the upper triangle. For
instance, the RS2 model lives on the diagonal, the model proposed in
\cite{Dubovsky:2001fj} lives at the point B.

\begin{figure}
\begin{picture}(250,230)
(0,20) \includegraphics[height = 9cm, width=17cm]{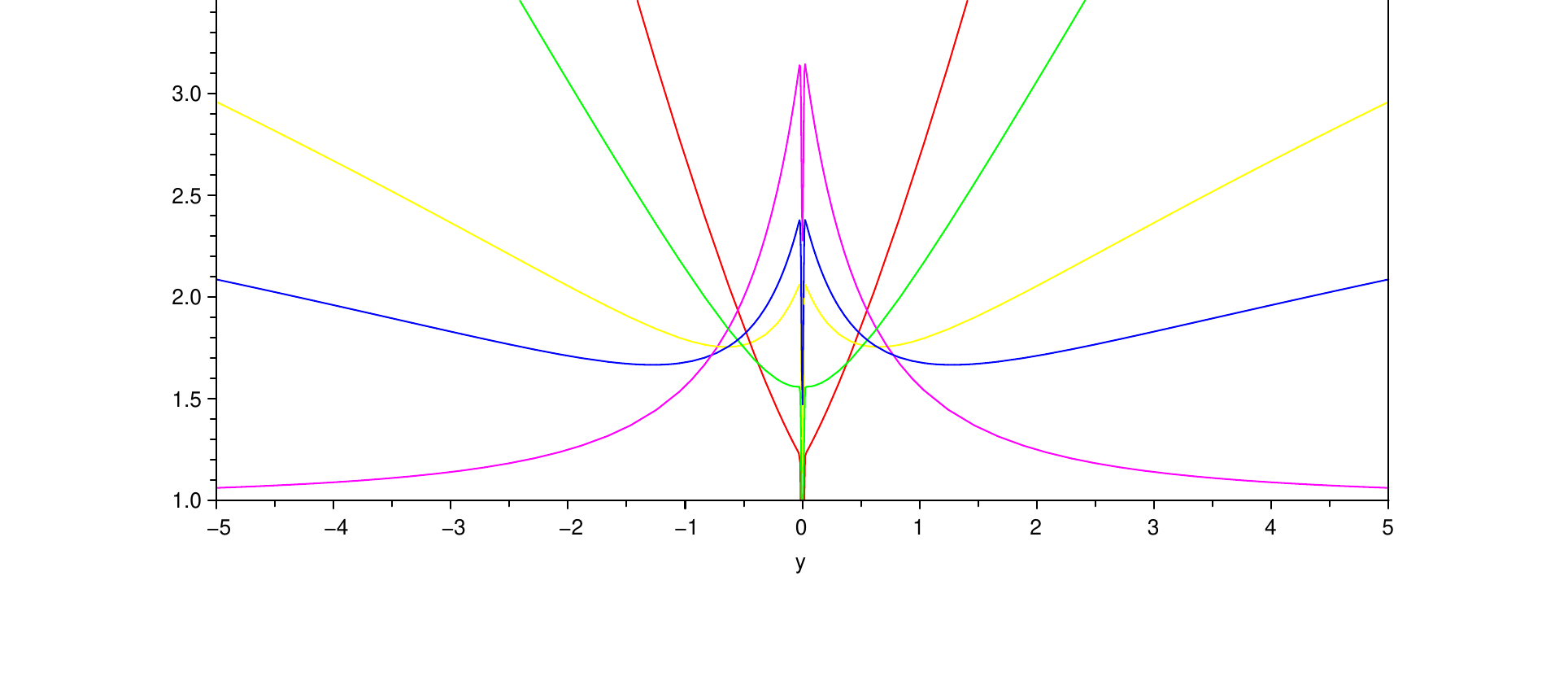}
\end{picture}
\caption{The effective potential for different values of $\zeta/\xi$. The
curves are ordered by steepness: for the steepest one the ratio
$\zeta/\xi$ is 0.3, then the values 0.5, 0.7, 0.8, 1 follow. The brane is
located at $y=0$ and $p=k=1$.}
\label{fig:generalpot}
\end{figure}

For the metric (\ref{eq:metricxizeta}) the potential in
(\ref{eq:SchroedingerEquation}), (\ref{eq:EnergyPotential}) takes the
following form
\begin{equation}
\label{eq:generalpotential}
V(y)=p^2\big(1+\xi k |y|\big)^{2(\zeta/\xi-1)}
+ \frac{(\xi^2+15\zeta^2) k^2}{4(1+\xi k
|y|\big)^{2}}-(\xi+3\zeta)k\delta(y)\,.
\end{equation}
We see that the Lorentz  violation contributes the first term of
the potential $V$, which is momentum-dependent. The ratio $\zeta/\xi$
determines the behavior of the potential at infinity. In our
considerations we restrict ourselves to the case with ``finite volume'' of
the fifth dimension\footnote{If the fifth dimension has an infinite volume
then there is  no in general  well defined four-dimensional effective
theory -- the model appears to be five-dimensional without any
localization properties on the brane. See some considerations in this
directions e.g. at \cite{Nelson:1999wu, Dvali:2000rv}.} with the following
condition satisfied
\[
\int\limits_{-\infty}^{+\infty}dy\,\sqrt{g}<\infty\,,
\]
which in our parametrization means that $\xi+3\zeta>0$. The corresponding
domain in \figref{fig:xizeta} has four regions I, II, III, and IV so let
us briefly discuss the form of the potential for all these regions. First,
we assume that both $\xi$ and $\zeta$ are nonnegative (regions I and II).
If $\zeta/\xi>1$ then the first term in (\ref{eq:generalpotential})
increases at large $y$ thereby producing infinite walls to the potential $V$.
Steepness of these walls is controlled by the same ratio $\zeta/\xi$. The
case with $\xi=\zeta$ corresponds to the Lorentz invariant background of
the RS2 model with $AdS_5$ as a bulk space. The case with
$\zeta/\xi<1$ does not differ from the RS2 model qualitatively --- the
potential has the same kind of a shape. In the region III the potential
(\ref{eq:generalpotential}) has poles at $|y|=-(\xi k)^{-1}$ but they are not physical, our coordinate system is not appropriate for this region. However, upon choosing a proper coordinates one can proceed with this region as well and observe similar behavior as in the region I. As far as we can conclude from (\ref{eq:generalpotential}), models located in the region IV
do not qualitatively differ from models of type II, indeed, the potential
has similar asymptotic behavior at infinity. At \figref{fig:generalpot} we
depicted the potential (\ref{eq:generalpotential}) at different positive
values of the ratio $\zeta/\xi$.

Employing the considerations from the beginning of this section we can
speculate on character of the perturbation spectra of type I and type II
models. A full exact solution for generic value of $\xi/\zeta$ is however
not possible, nevertheless approximate analysis leads to desired
understanding of the spectra of these models. Analogously with the
considerations from the previous subsection we can straightforwardly state
that the models of type I have a box-type potential and hence have
\textit{discrete} spectrum of the scalar perturbations, whereas the models
of type II have decaying potential at infinity and plane wave excitations
propagating along $y$-axis. The excitation spectrum of type II models is
therefore \textit{continuous} in general. Due to the delta-well it may
contain a zero mode as well. Properties of type II world were discussed in
\cite{Dubovsky:2001fj} (point B). Now we shall
investigate the scalar fluctuations in the completely different background
A with almost opposite properties.

\subsection{The Model A. Generic Solution and Static Potential}\label{sec:ModelA}

\begin{figure}
\begin{picture}(150,150)
(-30,15) \includegraphics[height = 8cm, width=15cm]{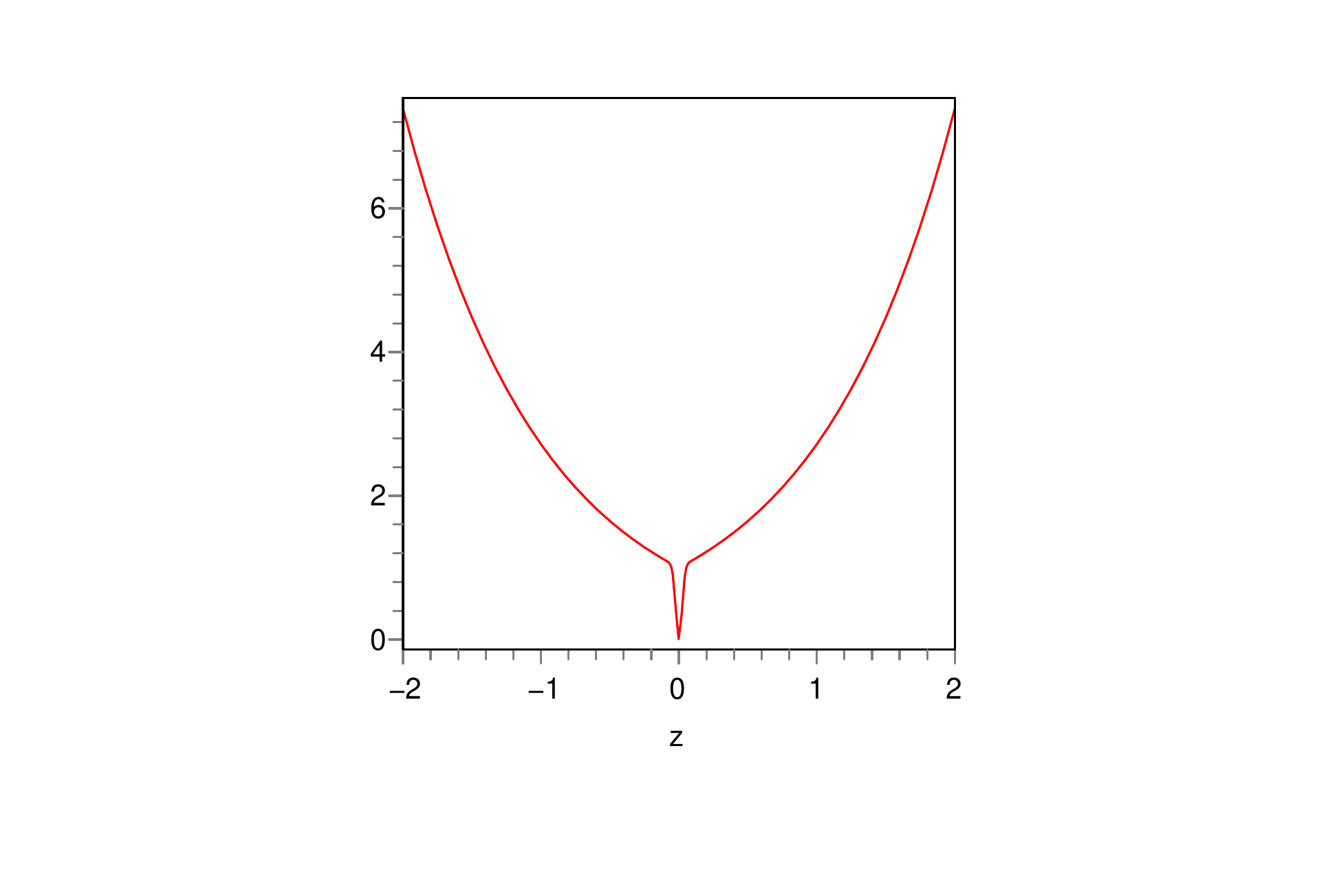}
\end{picture}
\caption{Shape of the model A potential. We put $p=k=1$ here.}
\label{fig:ourpotential}
\end{figure}

In order to proceed with the model A corresponding to the point $(0,1)$ in
\figref{fig:xizeta} one considers the Schr\"odinger equation in the
form (\ref{eq:KleinGordonEquationz}). The potential reads
\begin{equation}
\label{eq:ourpotential}
V(z) =
p^2\mathrm{e}^{2k|z|}+\frac{9}{4}k^2-3k\delta(z)\,.
\end{equation}
We see that the potential tends to infinity as $|z|\to\infty$, thus we
expect the perturbation spectrum to be discrete. Schematically the
potential \eqref{eq:ourpotential} is shown in \figref{fig:ourpotential}.
First, we solve (\ref{eq:KleinGordonEquationz}) with the
potential (\ref{eq:ourpotential}) for $z\neq 0$
\[
\chi''(z)+\left(E^{2}-p^2
\mathrm{e}^{2k|z|}-\frac{9}{4}k^2\right)\chi(z)= 0\,,
\]
then we impose the boundary conditions at the origin to take into account
the delta-function term
\begin{equation}
 \label{eq:MatchingA}
 [\chi'(z)]\Big\vert_0+3k\chi(0)= 0\,.
 \end{equation}
The normalizable solution is expressed via the following modified
 Bessel function
\begin{equation}
\label{eq:ScalarSolution}
\chi(z)=N\,\Mac_\nu\left(\frac{p}{k}\mathrm{e}^{k|z|}\right)\,,
\end{equation}
where $\nu=\sqrt{\frac{9}{4}-\frac{E^2}{k^2}}$ and $N$ is a normalization
constant which is determined by the  condition
\begin{equation}
\int\limits_{-\infty}^{+\infty}dz\,|\chi(z)|^2=1\,.
\label{Eq/Pg12/1:fluctpdf_0801}
\end{equation}
The matching condition (\ref{eq:MatchingA}) at $z=0$ yields
\begin{equation}
\label{eq:matchingsc}
\frac{p}{k}\frac{\Mac_{\nu+1}\left(\frac{p}{k}\right)}{\Mac_{\nu}\left(\frac{p}{k}\right)}=\frac{3}{2}+\nu\,.
\end{equation}
This equation relates the energy and momentum of  modes,
therefore it represents the dispersion relation. We consider it below.

\paragraph{Green Function.}

Let us now  find the brane-to-bulk Green function for the scalar field
$\phi$. In the momentum space it satisfies the following
equation\footnote{Due to $SO(3)$ rotational brane symmetry the Green
function depends only on absolute value of momentum $p$.}
\begin{equation}
\label{eq:EquationGreenFunction}
\left[-\partial_z^2-E^2+p^2
\mathrm{e}^{2k|z|}+\frac{9}{4}k^2-3k\delta(z)\right]\Delta_p(E,p,z)=\delta(z)\,.
\end{equation}
Using (\ref{eq:ScalarSolution}) one obtains the decaying solution
to this equation outside of the brane
\[
\Delta_p(E,p,z)=C(E,p)\Mac_{\nu}\left(\frac{p}{k}\mathrm{e}^{k|z|}\right)\,,
\]
where $C(E,p)$ is yet to be determined constant. Due to the presence of
the delta-functions in (\ref{eq:EquationGreenFunction}) one imposes the
following constraint
\[
[\partial_z \Delta_p(E,p,0)]\Big\vert_0+3k\Delta_p (E,p,0) = -1\,.
\]
This equation determs the constant $C(E,p)$. Now we are
ready to find the propagator. Let us put $z=0$ in order to extract the
brane-to-brane propagator
\[
\Delta_p(E,p,0)=\frac{2}{k}\left[\frac{p}{k}\frac{\Mac_{\nu+1}\left(\frac{p}{k}\right)}{\Mac_{\nu}\left(\frac{p}{k}\right)}-\nu-\frac{3}{2}\right]^{-1}\,.
\]
We see that the condition for the brane-to-brane propagator to have a pole
exactly coincides with (\ref{eq:matchingsc}). Later we  show that
(\ref{eq:matchingsc}) has only real valued solutions in terms of $E$,
hence all perturbation modes in our model are stable. In order to find the
static brane-to-brane Green function we put $E=0$ and obtain it in the
momentum space
\begin{equation}
\label{eq:potentialfourier}
G_p(0)\equiv\Delta_p(0,p,0)=\frac{1}{2k}\left[\frac{p}{k}
\frac{\Mac_{\sfrac{5}{2}}\left(\frac{p}{k}\right)}{\Mac_{\sfrac{3}{2}}
\left(\frac{p}{k}\right)}-3\right]^{-1}=\frac{k}{2p^2}+\frac{1}{2p}\,.
\end{equation}
Inverted Fourier transform leads us to the Green function in the
coordinate representation which represents the static potential
\begin{eqnarray}
\label{eq:GreenFunction}
G(r)\eq \int\frac{d^3
p}{(2\pi)^3}G_p(0)\mathrm{e}^{i\textbf{pr}}=\frac{4\pi}{(2\pi)^3
r}\int\limits_0^{+\infty}dp\,\left(\frac{k}{p}+1\right)
\frac{\mathrm{e}^{ipr}-\mathrm{e}^{-ipr}}{2i}\nln
\eq\frac{1}{2\pi^2r}\left(k\int\limits_0^{+\infty}dp\,
\frac{\sin(pr)}{p}-2i\int\limits_{0}^{+\infty}dp\,\mathrm{e}^{ipr}\right)=
\frac{k}{4\pi r}\left(1+\frac{2}{\pi kr}\right)\,.
\end{eqnarray}
The obtained potential describes the Coulomb (Newton) law with small
correction at large distances compared with the curvature scale $1/k$.
At small distances the interaction becomes five-dimensional. Note that
unlike the RS2 model \cite{Randall:1999vf} the correction to the potential
has a different power: $r^{-2}$ instead of $r^{-3}$ for the RS2 model.

\subsection{Spectrum of the model A}

\begin{figure}
\begin{center}
\includegraphics[height = 11cm, width=11cm]{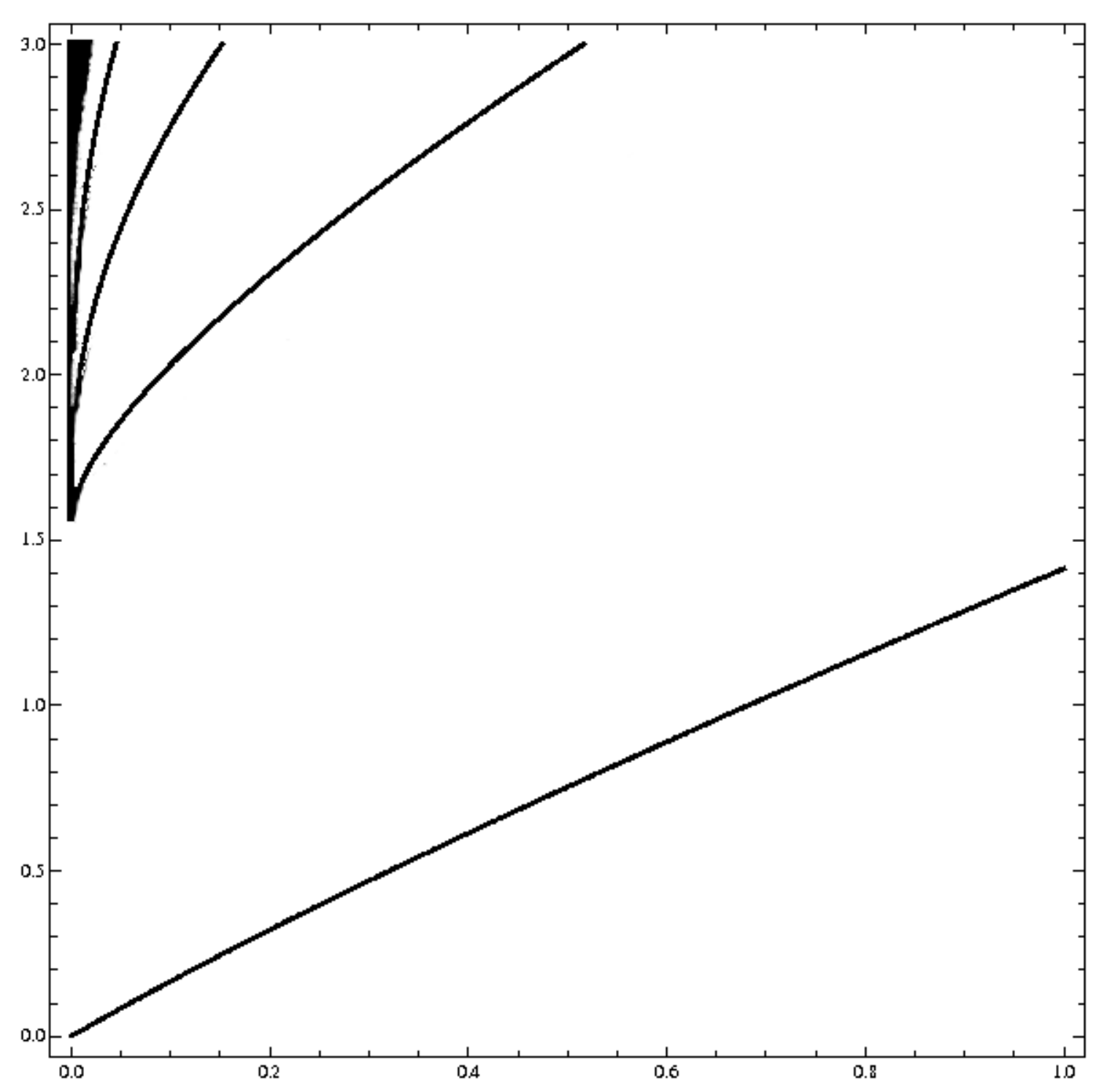}
\caption{Spectrum of the field $\phi$. The energy $E$ in the vertical
axis and the momentum $p$ in the horizontal axis are scaled in the units
of $k$. The higher modes $n=1,2,\dots$ start from $E=\sfrac{3}{2}k$.}
\label{fig:SpectrumScalar}
\end{center}
\end{figure}

Let us now discuss the spectrum of the scalar sector in the model A. We
proceed with expanding (\ref{eq:matchingsc}) at small and large momenta
and derive dispersion relations for different excitation modes. The
spectrum of the scalar field $\phi$ including the zero mode and higher modes
is shown in \figref{fig:SpectrumScalar}.
We observe that there exists zero mode and higher excitations. The
behavior of the higher excitations at small momenta is of particular
interest, as all curves collapse to one point
$E=\sfrac{3}{2}k$ as $p\to 0$.

\paragraph{Zero mode at small momenta.}

From (\ref{eq:matchingsc}) one observes that there exists a solution with
$E(p)\to 0$ as $p\to 0$ which corresponds to the zero mode. The
corresponding dispersion relation can be found if one performs an
expansion in the vicinity of the point $E=0,\,p=0$, and after
straightforward calculation one obtains
\begin{equation}
\label{zerodisp}
E^2 = 3p^2\left(1-\frac{p}{k}+\mathcal{O}(p^2)\right)\,.
\end{equation}
We see that the zero mode dispersion relation is not Lorentz invariant.
However, the low energy theory is approximately
Lorentz invariant in the sense that the dispersion relation $E^2\approx
3p^2$ has the form of the ordinary relativistic formula with rescaled
speed of light. From  (\ref{zerodisp}) one finds the speed of
propagation of scalar particles (group velocity $\partial E/\partial p$)
which is equal to $\sqrt{3}$ and exceeds a speed of tightly bounded
massless particles on the brane being equal to unity. It also exceeds the
speed of scalar particles considered in \cite{Dubovsky:2001fj} which is
always less than 1. Therefore, one can expect that while
moving in the parameter space in \figref{fig:xizeta} from point B to point
A throughout the upper right region, the speed of scalar particles will
increase from the values found in \cite{Dubovsky:2001fj} to $\sqrt{3}$
derived here.

Let us calculate the zero mode contribution to the static brane
Green function (\ref{eq:potentialfourier})
\begin{equation}
\label{eq:GreenZero}
G^{(0)}_p(0)=\frac{|\chi^{(0)}
(0)|^2}{E^2}\,.
\end{equation}
In order to find it one should know the zero mode normalization constant $N_{0}$
which can be determined at low momenta in the following way. One notes
that $\partial \Mac_{\nu }(x)/\partial \nu =\Mac_\nu (x)(\log x
+\mathcal{O}(1))$ (see, e.g.~\cite{Bateman:1953}),  so the zero mode wave
function can be represented as
\begin{equation}
\chi
^{(0)}(z)=N_{0}\Mac_{\frac{3}{2}}\left(\frac{p}{k}\mathrm{e}^{k|z|} \right)
\left[1+\mathcal{O}\left( p^{2}\log \frac{p}{k}\mathrm{e}^{k|z|}\right)
\right].
\label{Eq/Pg14/1:fluctpdf0611}
\end{equation}
After that the normalization constant can be calculated from
(\ref{Eq/Pg12/1:fluctpdf_0801})
\[
N_{0}^{2}=k \frac{3}{\pi }\left(\frac{k}{p}   \right)^{3}\left[1+\mathcal{
O}\left(\frac{p^{2}}{k^{2}}\log\frac{p}{k}   \right)   \right].
\]
Given $N_{0}$ and substituting the zero mode dispersion relation (\ref{zerodisp}) together with (\ref{Eq/Pg14/1:fluctpdf0611}) into (\ref{eq:GreenZero}) one finds the zero mode contribution to the static
Green function
\begin{equation}
G^{(0)}_p(0)=\frac{k}{2p^{2}}+\frac{1}{2p}+\mathcal{ O}(\log p)\,.
\label{Eq/Pg14/2:fluctpdf0611}
\end{equation}
We see that at low momenta the Green function (\ref{eq:GreenFunction}) is
saturated by the zero mode. The corrections $\mathcal{O}(\log p)$ emerging
in (\ref{Eq/Pg14/2:fluctpdf0611}) should be canceled out by contributions
from higher modes.

\paragraph{Higher modes at small momenta.}

As was mentioned above the mode equation (\ref{eq:KleinGordonEquationz})
can be written as a Schr\"{o}dinger equation with  a potential dependent
on a particle momentum. We can apply WKB method here for
understanding the spectrum of higher excitations\footnote{In order to find
the dispersion relation for the higher modes one can solve
(\ref{eq:matchingsc}) at the low momenta directly. However, we do not
perform this analysis here. For the fermionic sector, where the spectrum
of higher modes will appear to be very much similar to the scalar one
discussed here, we shall expand an analog of (\ref{eq:matchingsc})
directly.}. The quantization law reads\footnote{In our calculations we ignore $\delta$-function contribution to the potential $V$ as it  brings only small correction to the integral in the l.h.s of the WKB formula. Therefore we can also neglect $\half$ in the r.h.s. as it is irrelevant for our
precision.}
\begin{equation}
\label{eq:WKBlaw}
\oint \sqrt{E^2_n-V(z)}\, dz = 2\pi n\,,
\end{equation}
where the potential $V(z)$ is given in (\ref{eq:ourpotential}). Because of
$\mathbb{Z}_2$ reflection symmetry of the potential we have
\begin{equation}
2\cdot 2\int\limits_{0}^{z_n}\sqrt{E^2_n-V(z)}dz=2\pi
n\,,
\label{Eq/Pg17/1:fluctpdf_0801}
\end{equation}
where the classical turning points have the following coordinates for each
$n$
\[
z_n=\pm\frac{1}{2k}\log\frac{E^2_n-\frac{9}{4}k^2}{p^2}\,.
\]
Calculating integral in the l.h.s. of (\ref{Eq/Pg17/1:fluctpdf_0801}) one
has
\[
-\sqrt{E^2_n-\sfrac{9}{4}k^2-p^2}+\sqrt{E^2_n-\sfrac{9}{4}k^2}\,
\mbox{Arctanh}
\displaystyle{\sqrt{1-\frac{p^2}{E^2_n-\sfrac{9}{4}k^2}}}=\frac{\pi k}{2}
n\,.
\]
Expansion for $0<p\ll k$ yields
\[
-\sqrt{E^2_n-\sfrac{9}{4}k^2}\log{\frac{p}{k}}=\frac{\pi k}{2}n\,,
\]
or
\begin{equation}
\label{eq:WKBSpectum}
E^2_n = \frac{9}{4}k^2+\frac{\pi^2 k^2
n^2}{4\log^2\frac{p}{k}} \,.
\end{equation}
One can show that the WKB method used here is applicable provided that
$p/k < \mathrm{e}^{-n}$.

The obtained result (\ref{eq:WKBSpectum}) becomes more transparent if one
notes that at small momenta the potential (\ref{eq:ourpotential}) can be
approximated by the box-type potential
\[
p^{2}\mathrm{e}^{2k|z|}\simeq\left[
\begin{array}{cc}
0&\mbox{if } |z|<-\displaystyle\frac{1}{k}\log\frac{p}{k}\\
\\
\infty &\mbox{if } |z|>-\displaystyle\frac{1}{k}\log\frac{p}{k}
\end{array}
  \right.\,.
\]
Then the dispersion relation (\ref{eq:WKBSpectum}) immediately follows.
Note also, that if $p= 0$ the spectrum becomes continuous what corresponds
to a  particle which can propagate to infinity in complete agreement
with the results of \secref{sec:LIVSummary}.

\paragraph{Large momenta.}

At large momenta $p\gg k$ one can neglect the factor of $\frac{3}{2}$ in
the r.h.s. of (\ref{eq:matchingsc}) and   (\ref{eq:matchingsc}) becomes
equivalent to the following condition
\[
\Mac'_{\nu }\left(\frac{p}{k} \right)=0.
\]
This condition is considered in \appref{sec:Derivation} in the context
of fermionic spectrum at large momenta (see (\ref{eq:Pg28/3})). The result
is
\[
E=p+\mathcal{O}(p^{\frac{1}{3}})\,.
\]
Thus at large momenta, as expected, the theory becomes Lorentz invariant.

\section{Fermionic Fluctuations}\label{sec:Fermions}

In this section we discuss the spectrum of fermionic fluctuations in
braneworlds with broken Lorentz invariance. One can perform a qualitative
analysis for generic $(\xi,\zeta)$ configuration (\ref{eq:metricxizeta})
and reduce the appropriate five dimensional Dirac equations to a
Schr\"{o}dinger--type equations. Then, analogously to the scalar sector
discussed at \secref{sec:Scalars}, one can establish that the asymptotic
behavior of the metric at infinity is responsible for the actual pattern
of the spectrum. For instance, if $\xi\geq\zeta$ the spectrum is
continuous. In this section we refrain from considering qualitative
picture with generic ansatz (\ref{eq:metricxizeta}) but focus our
attention on the model A where exact solution is obtained and analyzed.

We consider the four-component spinor field $\Psi$ on the five-dimensional manifold with metric \eqref{eq:MetricOur}
\begin{equation}\label{eq:DiracMassTerm}
S=\int dt d\textbf{x}\int\limits_{-\infty}^{+\infty}dz\,\sqrt{g}
\left(i\bar{\Psi}\nabla\!\!\!\!\!\big/\,\, \Psi - m_{\psi}\bar{\Psi}\Psi\right)\,,
\end{equation}
where $m_{\psi}$ is the five-dimensional mass of the fermion which may however
depend on $z$. In this section we study two cases:

\textit{i)} $m_{\psi}=m\cdot\mbox{sign}(z)$. It corresponds to a usual
(Lorentz invariant) way of localization of fermions in extra
dimensional setups (see, e.g. \cite{Gherghetta:2000qt,
Grossman:1999ra,Gherghetta:2006ha} and for more recent treatments \cite{Liu:2008pi}). It is worth noting that in this case
the action (\ref{eq:DiracMassTerm}) is invariant under $\Integers_{2}$
reflection symmetry $z\to -z$ if the fermion field transforms as
\begin{equation}
\label{Eq/Pg16/1A:fluctpdf0611}
\Psi (-z)=\mp \gamma _{5} \Psi (z),
\end{equation}
where phase $\mp$ can be only determined by fermion interactions.

\textit{ii) $m_{\psi }=m$.} In this case we deal with a usual massive
fermion and the action does not possess the $\Integers_{2}$
symmetry. It does not play a role, however, since this case has
only an illustrative character and we are interested in the limit $m\to 0$
which can be easily obtained in this case  and which restores
$\Integers_{2}$ symmetry.

\paragraph{Generic Solution.}

The Dirac equation obtained from (\ref{eq:DiracMassTerm}) reads
\begin{equation}
\label{eq:DiracEquationGen}
i\Gamma^A \nabla_A
\Psi(x,z)-m_{\psi}\Psi(x,z)=0\,,
\end{equation}
where $\nabla_A = \dpod{A}+\omega_A$ is the covariant derivative with the
spin--connection
\[
\omega_{A}=
\begin{pmatrix}
0,&-\ihalf k\gamma_5\gamma_i \mathrm{e}^{-k|z|},&0
\end{pmatrix}
\,\text{sign}(z)\,.
\]
The five matrices $\Gamma^A$ ($A=0,1,2,3,5$) obeying the Clifford algebra relations
\[
\left\{\Gamma^A,\Gamma^B\right\}=2g^{AB}\,,
\]
are related to the $\gamma$-matrices ($\mu,\nu=0,1,2,3$)
\begin{equation}
\left\{\gamma^\mu,\gamma^\nu\right\}=2\eta^{\mu\nu},\quad \gamma^5=\gamma
_{5}=i\gamma^0\gamma^1\gamma^2\gamma^3\,,
\label{Eq/Pg16/1:fluctpdf0611}
\end{equation}
as follows
\[
\Gamma^0=\gamma^0,\quad \Gamma^i=\mathrm{e}^{k|z|}\gamma^i,\quad
\Gamma^5=-\Gamma _{5}=-i\gamma^5\,.
\]
In order to simplify further calculations it is convenient to work with
the rescaled field
\begin{equation}
\label{eq:PsiRescaling}
\Psi=\mathrm{e}^{\sfrac{3}{2}k|z|}\psi\,,
\end{equation}
and apply the four-dimensional Fourier transform to the field $\psi$
\[
\psi(t,\textbf{x},z)=\frac{1}{(2\pi)^2}\int
dEd\textbf{p}\,\mathrm{e}^{iEt-i\textbf{px}}\psi(E,\textbf{p},z)+\mbox{h.c.}\,.
\]
In this relation and in what follows we assume that $E>0$. Then due to
(\ref{eq:PsiRescaling}) the normalization condition for the field $\psi
(E,\mathbf{p},z)$ has the  canonical form
\begin{equation}
\int\limits_{-\infty}^{+\infty}dz \sqrt{g}\,\bar{\Psi }\gamma _{0}\Psi  =
\int\limits_{-\infty}^{+\infty}dz\,\psi ^{\dag}(E,\mathbf{p},z)\psi
(E,\mathbf{p},z)=1\,.
\label{eq:Pg15/1}
\end{equation}
In terms of the new field the Dirac equation (\ref{eq:DiracEquationGen})
takes the  form
\begin{equation}
\label{eq:DiracEquationMom}
\left(E\gamma^0-\mathrm{e}^{k|z|}\gamma^3
p-\gamma^5\dpod{z}\right)\psi+m_{\psi }\psi=0\,.
\end{equation}
where we set $\textbf{p}=(0,0,p)$ for simplicity. In order to solve this
equation we use the following decomposition of the spinor
\begin{equation}
\psi =\sum \limits_{\alpha ,\beta }^{}\phi _{\alpha, \beta
}(E,p,z)U_{\alpha, \beta },
\label{Eq/Pg17/1:fluctpdf0611}
\end{equation}
where $\alpha $ and $\beta $ independently take values $\pm$, $\phi
_{\alpha, \beta }$ are four unknown functions, and $U_{\alpha, \beta }$
are four constant spinors with the following properties
\begin{eqnarray}
2S_{3}U_{\alpha ,\beta }&\equiv&i\gamma_1\gamma_2U_{\alpha
,\beta }=\alpha U_{\alpha ,\beta }\nonumber\\
\gamma _{0}U_{\alpha ,\beta }&=&\alpha \beta U_{\alpha ,\beta
}\nonumber\\
\gamma _{3}U_{\alpha ,\beta }&=& \beta U_{\alpha ,-\beta }\nonumber\\
\gamma _{5}U_{\alpha ,\beta }&=& -U_{\alpha ,-\beta }\,.
\label{Eqn/Pg18/1:fluctpdf0611}
\end{eqnarray}
Let us say a few words concerning the above relations. Because of
$S_{3}$ (the third component of the spin) commutes with
(\ref{eq:DiracEquationMom}) it is easy to use the basis composed of the
eigenvectors of $S_{3}$ --- this is what the first property
tells us. Since $\gamma_0$ commutes with $S_{3}$ one can
choose the basis in which the eigenvectors of $S_{3}$ are
eigenvectors of $\gamma_0$ as well --- this is what the second property
tells us. The first two properties correspond to fixing of the orthogonal
basis up to four unknown constants. Since $\gamma_3$ commutes with
$S_{3}$ and anticommutes with $\gamma_0$ one can check that
$\gamma_3 U_{\alpha ,\beta }\sim U_{\alpha ,-\beta }$ --- the third
property fixes the constant in this relation and, thereby, fixes the basis
up to two constants which can be found (up to phases) from the
normalization conditions
\[
U^{\dag }_{\alpha ',\beta '}U_{\alpha ,\beta }=2\delta _{\alpha '\alpha
}\delta _{\beta '\beta }\,.
\]
The last property follows directly from the first three properties and
from the definition of $\gamma _{5}$ \eqref{Eq/Pg16/1:fluctpdf0611}.

Substituting the expansion \eqref{Eq/Pg17/1:fluctpdf0611} into
\eqref{eq:DiracEquationMom} one obtains the following equation
\begin{equation}
(\partial _{z}-\beta p\mathrm{e}^{k|z|})\phi _{\alpha ,\beta }=(\alpha
\beta E-m_{\psi })\phi _{\alpha ,-\beta }\,.
\label{Eq/Pg18/1:fluctpdf0611}
\end{equation}
Acting on both sides of this equation by the operator $\partial
_{z}+\beta p\mathrm{e}^{kz}$ and using \eqref{Eq/Pg18/1:fluctpdf0611} with
$\beta \to -\beta $ one ends up with the  second order equation
\begin{equation}
\label{Eq/Pg18/2:fluctpdf0611}
\left(\partial ^{2}_{z}+E^2-m^2-p^2
\mathrm{e}^{2k|z|} -\beta pk\,\mbox{sign}(z) \mathrm{e}^{k|z|}\right)\phi
_{\alpha ,\beta }=-2m\delta (z)\phi _{\alpha ,-\beta }\,,
\end{equation}
where we assume that $m_\psi=m\cdot\mbox{sign}(z)$. It is easy to check
that the normalizable solution of \eqref{Eq/Pg18/2:fluctpdf0611} at
$z>0,\,p>0$ has the  form\footnote{No summation over
repetitive indexes $\alpha $ and $\beta$ is assumed.}
\begin{equation}
\label{Eq/Pg19/1:fluctpdf0611}
\phi^>_{\alpha ,\beta}(z)=C^{>}_{\alpha
,\beta }\xi _{\beta }\left(\frac{p}{k}\mathrm{e}^{k|z|} \right)\,,
\end{equation}
where
\begin{equation}
\label{eq:Pg17/1}
\xi_{\beta }(x)=\sqrt{x}\left[\Mac_{\nu
+\frac{1}{2}}(x)-\beta \Mac_{\nu-\frac{1}{2}}(x)\right]\,.
\end{equation}
In what follows we use the superscripts ``$>$'' for the functions at $z>0$
and ``$<$'' for the functions at $z<0$, $C_{\alpha ,\beta }^{>,<}$ are
constants, and $\nu=k^{-1}\sqrt{m^2-E^2}$. It is worth noting that the
Dirac operator in the problem under consideration is Hermitian. Thus  the
energy $E$ has to be real (in fact it follows directly from the dispersion
relation). It means in particular that $\nu $ has to be either
real ($E<m$) or pure imaginary ($E>m$). In both cases the sign of
$\nu $ does not play any role: it can be absorbed into the definitions of
$C_{\alpha ,\beta }$ because of $\xi _{\beta }(-\nu )=-\beta \xi _{\beta
}(\nu )$. So  we assume that either real ($E<m$) or imaginary ($E>m$) part
of $\nu$ is positive.

One can see from (\ref{Eq/Pg18/2:fluctpdf0611}) that under the reflection
transformation $z\to -z$ only the last term in the l.h.s. changes the
sign. Thus  the solution at $z<0$ can be represented as
\[
\phi ^{<}_{\alpha ,\beta }(z)=C^{<}_{\alpha ,\beta }\xi _{-\beta }
\left(\frac{p}{k}\mathrm{e}^{k|z|} \right)\,.
\]
It follows from \eqref{Eq/Pg18/2:fluctpdf0611} that the solutions for
$p<0$ can be obtained from \eqref{Eq/Pg19/1:fluctpdf0611} and
\eqref{eq:Pg17/1} by interchanging $\beta \to -\beta $. The eight
unknown constants $C_{\alpha ,\beta }^{>,<}$ have to be determined from
the normalization condition \eqref{eq:Pg15/1}, from the continuity
conditions at the origin
\begin{equation}
\label{eq:Pg18/2}
\phi_{\alpha ,\beta }  ^{>}(0)=\phi _{\alpha ,\beta }
^{<}(0)\,,
\end{equation}
and from the matching conditions for the derivatives which follow from
(\ref{Eq/Pg18/2:fluctpdf0611})
\begin{equation}
\label{eq:Junction}
\partial_z \phi_{\alpha ,\beta } ^>(0)-\partial_z \phi
_{\alpha ,\beta }^< (0) = -2m\phi _{\alpha ,-\beta }(0)\,.
\end{equation}
At fixed $\alpha $, \eqref{eq:Pg18/2} and \eqref{eq:Junction} form a
set of four linear homogeneous equations for the four unknown constants
$C_{\alpha ,\beta }^{>,<}$. Hence in order this set to be self-consistent its determinant
should be equal to zero. Calculating the determinant we end
up with the  equation
\begin{equation}
\label{eq:ChiMatching}
t^{2}\left(\frac{\partial }{\partial t}(\xi
_{+}(t)\xi _{-}(t))\right)^{2}=\frac{4m^{2}}{k^{2}}(\xi _{+}\xi
_{-})^{2}\,,
\end{equation}
where $t=p/k$. This equation represents the dispersion relation $E=E(p)$
for the fermion field in question.

The equations (\ref{eq:Junction}) and (\ref{eq:ChiMatching}) can be
considerably simplified if one notes that the functions
$\xi_\beta$ satisfy the  relations
\[
\frac{d \xi _{\beta }(x)}{d x}=-\frac{\nu }{x}\xi _{-\beta }+\beta \xi
_{\beta }\,.
\]
Using these identities one obtains from (\ref{eq:Pg18/2}) and
(\ref{eq:Junction}) the relations for the constants $C_{\alpha
,\beta }^{>,<}$
\begin{eqnarray}
\label{Eqn/Pg20/1A:fluctpdf0611}
C^{<}_{\alpha ,\beta }\eq \frac{\xi
_{\beta }(t)}{\xi _{-\beta }(t)}C^{>}_{\alpha ,\beta }\,,\nln C_{\alpha
,-\beta }^{>}\eq\gamma \frac{\xi _{\beta }(t)}{\xi _{-\beta
}(t)}C^{>}_{\alpha ,\beta }\,,
\end{eqnarray}
where $\gamma =\pm 1$ corresponds to the two possible signs of the square
root of (\ref{eq:ChiMatching}). Making use of
(\ref{Eqn/Pg18/1:fluctpdf0611}) one verifies that $\gamma$ corresponds to
the $\mp$  signs in (\ref{Eq/Pg16/1A:fluctpdf0611}). Thus $\gamma$ remains
ambiguous and can be only fixed by the fermion interactions. As it
follows from (\ref{Eqn/Pg20/1A:fluctpdf0611}), at fixed $\alpha $ we have
only one unknown constant, say
\begin{equation}
\label{Eq/Pg21/1A:fluctpdf0611}
C_{\alpha }\equiv \xi_+ (t)C^{>}_{\alpha
,+ }=\gamma \xi _{-}(t)C^{>}_{\alpha ,-}\,,
\end{equation}
which is fixed by the normalization condition (\ref{eq:Pg15/1}). It
means that at fixed $\alpha $ and positive energy we have only one degree
of freedom --- a particle with the polarization $\alpha $.

It is worth noting here that $\nu $ cannot vanish at any value of energy.
Indeed, if $\nu = 0$ then $\xi _{+}\equiv 0$. It means that
$C_{\alpha ,-}\equiv 0$ (see (\ref{Eqn/Pg20/1A:fluctpdf0611})), and, so,
both $\phi _{\alpha ,\pm}\equiv 0$, i.e. there is no normalized solution
in this case.

Let us now substitute (\ref{Eqn/Pg20/1A:fluctpdf0611}) into
(\ref{Eq/Pg19/1:fluctpdf0611}) and afterwards into
(\ref{Eq/Pg18/1:fluctpdf0611}). Then one finds that
\begin{equation}
\label{Eq/Pg25/1A:fluctpdf0611}
\frac{\xi _{-\beta }(t)}{\xi _{\beta
}(t)}=\gamma \frac{\mu -\alpha\beta \varepsilon }{\nu }\,,
\end{equation}
where we have introduced the dimensionless energy $\varepsilon =E/k$ and
mass $\mu =m/k$. In fact this equation does not depend on $\beta $: if one
changes $\beta \to -\beta $ then by making use of the definition of $\nu$ in
the r.h.s of (\ref{Eq/Pg25/1A:fluctpdf0611}) one obtains the same
equation. Thus  (\ref{Eq/Pg25/1A:fluctpdf0611}) is equivalent to the
equation
\begin{equation}
\frac{\xi _{-}(t)}{\xi _{+}(t)}=\gamma \frac{\mu -\alpha \varepsilon }{\nu
}\,.
\label{Eq/Pg20/1A:fluctpdf0611}
\end{equation}
This relation is nothing but the dispersion
relation $E =E(p)$. One can check that (\ref{Eq/Pg20/1A:fluctpdf0611})
follows directly from (\ref{eq:ChiMatching}). However, from
(\ref{eq:ChiMatching}) one can obtain another branch of solutions which
corresponds to the substitution $\alpha \to -\alpha $ in
(\ref{Eq/Pg20/1A:fluctpdf0611}). This branch does not
satisfy (\ref{Eq/Pg18/1:fluctpdf0611}) and should be rejected.

Let us now consider the case $m_{\psi }=m$. In this case the term with the
delta-function in the r.h.s. of (\ref{Eq/Pg18/2:fluctpdf0611}) is absent.
Equations (\ref{Eq/Pg19/1:fluctpdf0611})---(\ref{eq:Pg18/2}) still
hold, in (\ref{eq:Junction}), (\ref{eq:ChiMatching}) the r.h.s. vanishes.
Thus the equation yielding the dispersion relation in this case (an analog
of (\ref{Eq/Pg20/1A:fluctpdf0611})) follows from  (\ref{eq:ChiMatching})
\begin{equation}
\label{Eq/Pg21/1:fluctpdf0611}
\frac{\xi _{-}(t)}{\xi _{+}(t)}=i\gamma\,,
\end{equation}
where $\gamma =\pm 1$. An analog of (\ref{Eqn/Pg20/1A:fluctpdf0611}) reads
as follows
\begin{eqnarray}
\label{Eqn/Pg21/1:fluctpdf0611}
C^{<}_{\alpha ,\beta }\eq \frac{\xi
_{\beta }(t)}{\xi _{-\beta }(t)}C^{>}_{\alpha ,\beta }=-i\beta \gamma
C^{>}_{\alpha ,\beta }\,,\nln C_{\alpha ,-\beta }^{>}\eq \frac{\nu }{\mu
-\alpha \beta \varepsilon }C^{>}_{\alpha ,\beta }\,.
\end{eqnarray}
The second equation in (\ref{Eqn/Pg21/1:fluctpdf0611}) follows from
(\ref{Eq/Pg18/1:fluctpdf0611}). This equation in fact does not depend on
$\beta $ --- interchanging $\beta\to -\beta  $ leads to the same equation
if one takes into account the definition of $\nu $. Thus  we have no other
constraints on the constants $C_{\alpha ,\beta }$ and at fixed $\alpha $
we have only one degree of freedom.

Let us now discuss the spectrum of the fields $\phi_{\alpha ,\beta }$
which follows from (\ref{Eq/Pg20/1A:fluctpdf0611}) and
(\ref{Eq/Pg21/1:fluctpdf0611}).

\subsection*{Spectrum of the model A}

\begin{figure}
\begin{center}
\vspace{-1cm}
\includegraphics[height = 9cm, width=9cm]{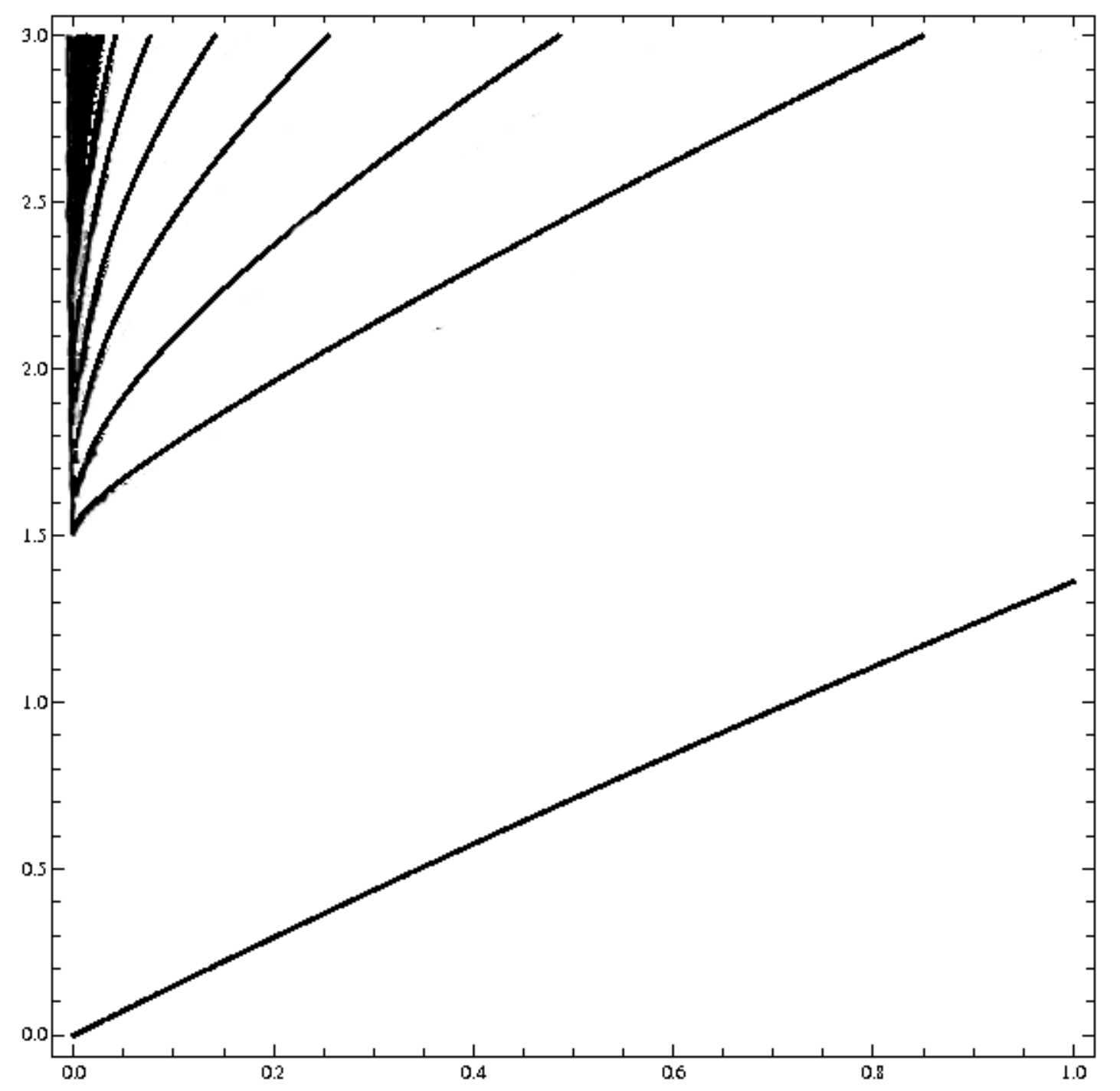}
\vspace{1.5cm} \vfill \hspace{-1mm}
\includegraphics[height = 9cm, width=9.3cm]{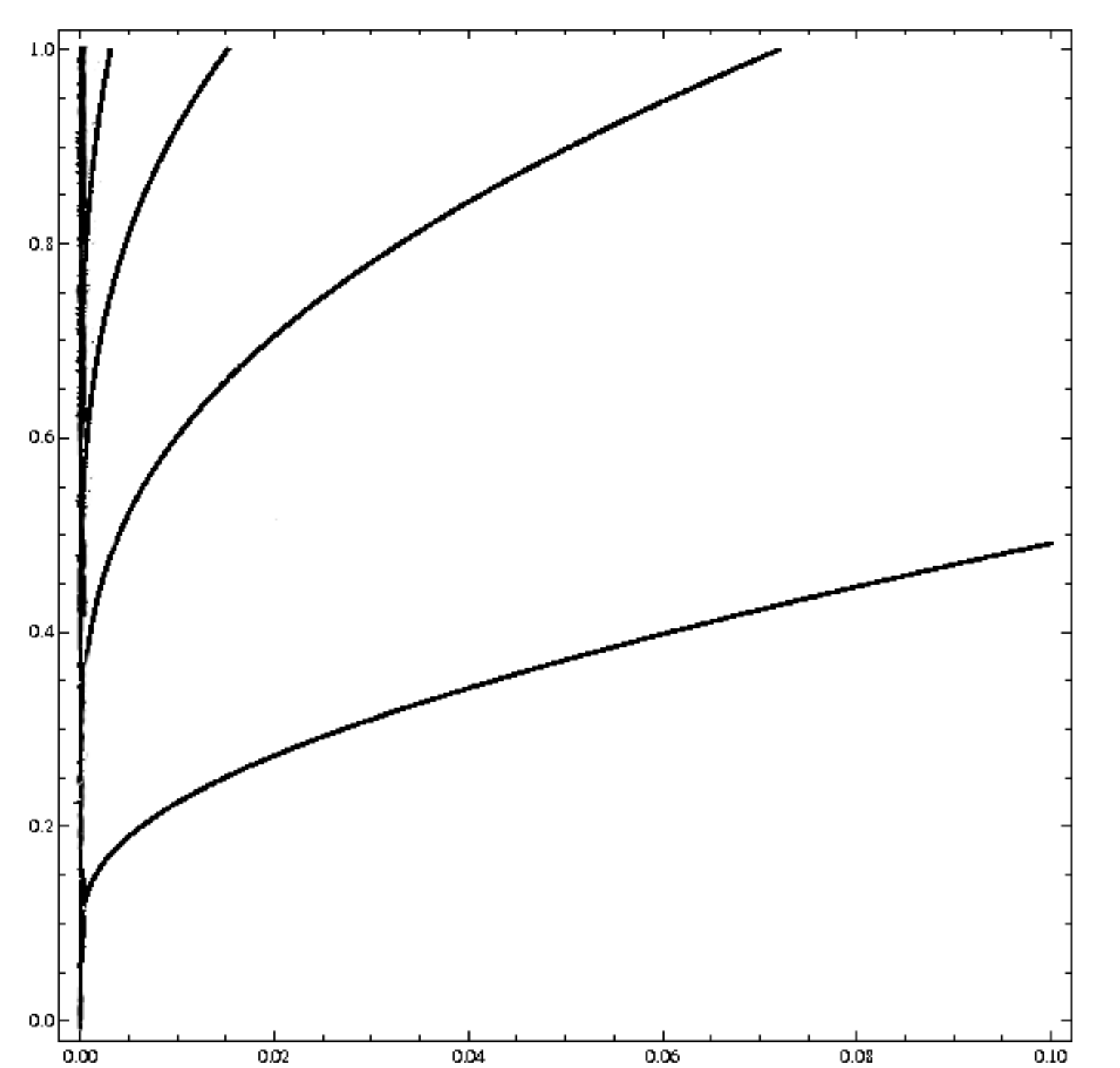}
\caption{Fermionic spectra $\varepsilon =\varepsilon (t)$ for $\mu
=\sfrac{3}{2} $, $\gamma =1$ and $m_{\psi }=m\cdot\mbox{sign}(z)$ (upper
plot), and $\mu =0$ (lower plot). Straight line starting from the
origin of the upper plot shows the zero mode dispersion relation. Other
curves originating from the point $(0,\sfrac{3}{2})$ show
dispersion laws for higher modes at different $n=1,2,\dots$. At the lower plot all curves
start from the origin.}
\label{fig:Spectrum}
\end{center}
\end{figure}

Here we give a summary of the fermionic spectrum of the model A at low and
high momenta and at different values of the bulk mass. The detailed
calculations are given in \appref{sec:Derivation}. The spectrum pattern at
low momenta is shown in \figref{fig:Spectrum}. We begin with the case $m
_{\psi }=m\cdot \mbox{sign}(z)$.

\paragraph{Zero mode at low momenta.}

A zero or gapless mode is a state with energy $E(p)\to 0$ as momentum
$p\to 0$. From the four-dimensional point of view a zero mode corresponds
to a massless state. As we will see the zero mode is present in the
fermionic spectrum of the model both in the case $m=0$ and $m\neq0$.
However, the dependence of the energy on the momentum $p \ll m,k$ is
different for different values of the bulk mass $m$. Here we consider only
the case of nonzero mass. As is shown in \appref{sec:Derivation}, the
dispersion relation has the following form
\begin{equation}
\label{eq:FermionZeroDispGeneral}
E=\left(\frac{2m}{2m-k}\right)p+2m
\frac{\Gamma\left(\half-\sfrac{m}{k}\right)}
{\Gamma\left(\half+\sfrac{m}{k}\right)}
\left(\frac{p}{2k}\right)^{\frac{2m}{k}}+
\mathcal{O}\left(p\log{\frac{p}{k}}\right)\,.
\end{equation}
From this formula the following three regimes can be derived
\begin{itemize}
\item At $m> k/2$ the dispersion relation is
\begin{equation}
\label{eq:Pg20/1A}
E\simeq \left(\frac{2m}{2m-k}\right)p\,.
\end{equation}
We see that in the limit $m\gg k$ the dispersion relation becomes Lorentz
invariant as it should be as this limit corresponds to the flat case.

\item At $m=k/2$ we have
\begin{equation}\label{eq:Pg20/1B}
E\simeq -p\log\frac{p\,\mathrm{e}^{\gamma_{E}}}{2k}\,,
\end{equation}
where $\gamma_{E}\simeq 0.577$ is Euler constant.

\item At $0<m<k/2$ we have
\begin{equation}\label{eq:Pg20/1C}
E\simeq 2m\frac{\Gamma\left(\frac{1}{2}-\frac{m}{k}\right)}
{\Gamma\left(\frac{1}{2}+\frac{m}{k}\right)}\left(\frac{p}{2k}
\right)^{\frac{2m}{k}}\,.
\end{equation}
\end{itemize}
In all these regimes it turns out that $\alpha=-1$,
and $\gamma =1$. Thus  the zero mode describes a massless polarized
particle with helicity $-1$.

\paragraph{Higher modes at low momenta.}

We start with the case $m_{\psi}=m\cdot\mbox{sign}(z)$. Similarly to the scalar spectrum discussed in \secref{sec:Scalars} one
can study the spectrum of higher fermionic modes employing the WKB
approximation. Here we, however, present the dispersion relations for the
higher modes obtained directly from (\ref{Eq/Pg20/1A:fluctpdf0611}).
Solving this equation at $p\ll m,k$ one finds (the detailed calculations
are given in \appref{sec:Derivation})
\begin{equation}\label{eq:Pg20/1}
E_n = m\sqrt{1+\left(\frac{\pi
nk}{2\Psi\left(\half\right)m-\gamma k -2m\log\frac{p}{2k}}\right)^{2}}\simeq
m\sqrt{1+\left(\frac{\pi nk}{2m\log\frac{p}{k}}\right)^{2}}\,,
\end{equation}
where $\Psi(z)$ is Euler digamma function and $n\in\mathbb{N}$ is the mode
number. The helicity depends on the mode number as $\alpha=(-1)^{n+1}$.
The latter approximate equality in (\ref{eq:Pg20/1}) is obtained by the WKB
approximation and is similar to (\ref{eq:WKBSpectum}) for the scalar case.
It is worth noting that the first equality in (\ref{eq:Pg20/1}) is valid
only for $n$ such that the term in the parenthesis is less than unity.
At $n$ large enough the character of the spectrum does not
change: it remains to be discrete and $E\to m$ as $p\to 0$.

It follows from (\ref{eq:Pg20/1}) that at fixed level number $n$
there are two states with different energies\footnote{It worth to note
that $\gamma $ (sign in (\ref{Eq/Pg16/1A:fluctpdf0611})) plays a role
of the parameter of the model. Thus for the fixed $\gamma $ there is only
one state.} $E(\gamma =1)$ and $E(\gamma =-1)$. Moreover, these two massive
states are polarized and have the same helicity $\alpha =(-1)^{n+1}$. So
the spectrum of the model for the positive energies is completely
non-degenerate.

Now we consider the case with the constant bulk mass $m_{\psi
}=m$ at small momentum $p\ll m,k$. In this case we obtain the following
dispersion relation
\begin{equation}
\label{Eqn/Pg24/1:fluctpdf0611}
E_n=\sqrt{m^{2}+\left(\frac{k\pi(2n+1)
}{4\Psi\left(\half\right) -4\log\frac{p}{2k}}\right)^{2}}\,,
\end{equation}
where $n$ is a nonnegative integer number. We also have
$\gamma=(-1)^{n+1}$. It is worth noting that $\gamma $ rather then $\alpha
$ is fixed in this case and $E_{n}$ does not depend on $\alpha $ at all.
Thus for fixed $n$ the spectrum is doubly degenerate -- there are two
modes with helicities $\alpha =\pm 1$. One also sees that there is a
smooth limit $m\to 0$ in this case in which all modes become
gapless.

\paragraph{Modes at high momenta.}

At momenta large enough, $p\gg m,k$, the dispersion relations for all
modes in the both cases $m_{\psi }=m\cdot\mbox{sign}(z)$ and $m_{\psi }=m$
become
\begin{equation}
\label{eq:Pg20/2}
E=p+\mathcal{ O}(p^{\sfrac{1}{3}})\,,
\end{equation}
and the correction in (\ref{eq:Pg20/2}) is
positive. Thus  at large momenta the Lorentz invariant dispersion relation
is restored.

\paragraph{Chiral properties of solutions.}

\begin{figure}[h!]
\includegraphics[height = 12cm, width=15cm]{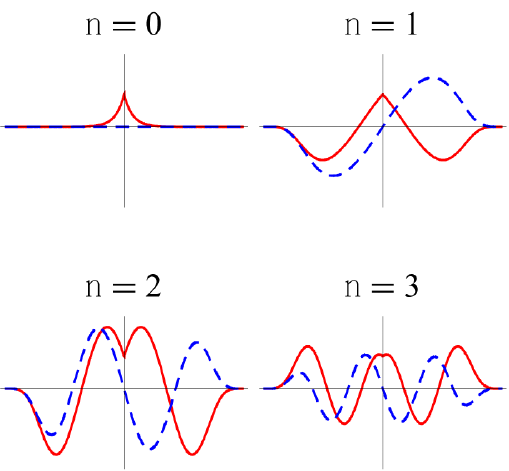}
\caption{Mode profiles at $n=0,1,2,3$ for $m=2k,\,p=0.01$. Red solid curves
correspond to left-handed and dashed blue curves correspond to
right-handed spinors. On the upper left plot (n=0) the right-handed
component is strongly suppressed.}
\label{fig:ChiralPlot}
\end{figure}

It is a standard lore that in the Lorentz invariant case for $m_{\psi
}=m\cdot\mbox{sign}(z)$ a zero mode appears in the spectrum and this zero
mode is chiral from the four-dimensional point of view. Let us consider
what happens in our case. Using (\ref{Eqn/Pg18/1:fluctpdf0611}) one
finds
\[
\psi _{L,R}=\frac{1\mp\gamma_{5} }{2} \psi =\frac{1}{2}\sum
\limits_{\alpha ,\beta }^{}U_{\alpha ,\beta }(\phi _{\alpha ,\beta
}\pm\phi _{\alpha ,-\beta })\,.
\]
From this equation by means of (\ref{Eq/Pg19/1:fluctpdf0611}) and
(\ref{Eq/Pg21/1A:fluctpdf0611}) one has
\begin{equation}
\label{Eq/Pg25/1:fluctpdf0611}
\psi^{>} _{L,R}=\left(\frac{\xi
_{+}(t\mathrm{e}^{k|z|})}{\xi _{+}(t)}\pm\gamma \frac{\xi
_{-}(t\mathrm{e}^{k|z|})}{\xi _{-}(t)}\right)\frac{1}{2}\sum
\limits_{\alpha }^{}C_{\alpha }(U_{\alpha ,+}\pm U_{\alpha ,-})\,.
\end{equation}
The corresponding expression at  $z<0$ can be obtained from
(\ref{Eq/Pg25/1:fluctpdf0611}) by interchanging $\xi
_{-}\leftrightarrow\xi _{+}$. Thus both left-handed and right-handed
components do not vanish as functions of $z$. In particular, the zero mode
contains both left-handed and right-handed spinors. However, for the zero
mode $\gamma=1$ and the right-handed spinor vanishes on the brane $z=0$.
So, in this respect, one can say that only left-handed spinor is localized
on the brane. This statement is true for higher modes as well: if $\gamma
=1$ only left-handed spinors are localized on the brane, whereas,
right-handed spinors are localized for $\gamma =-1$. In both cases
depending on $\alpha$ the modes have different helicities (for the zero
mode $\alpha =-1$ and helicity is $-1$). In \figref{fig:ChiralPlot} some
mode profiles from \eqref{Eq/Pg25/1:fluctpdf0611} are shown.

In the case $m_{\psi }=m$ an analog of (\ref{Eq/Pg25/1:fluctpdf0611}) is
\begin{equation}\label{Eq/Pg25/2:fluctpdf0611}
\psi ^{>}_{L,R}=\sum \limits_{\alpha }^{} \left(\frac{\xi
_{+}(t\mathrm{e}^{k|z|})}{\xi _{+}(t)}\pm i\gamma \frac{\nu }{\mu -\alpha
\varepsilon }\frac{\xi _{-}(t\mathrm{e}^{k|z|})}{\xi
_{-}(t)}\right)\frac{C_{\alpha }}{2}[U_{\alpha ,+}\pm U_{\alpha ,-}]\,,
\end{equation}
so the chiral components do not vanish even on the brane. However, in the
massless limit the expression in the parentheses in
(\ref{Eq/Pg25/2:fluctpdf0611}) tends to
\[
\frac{\xi _{+}(t\mathrm{e}^{k|z|})}{\xi _{+}(t)}\pm \gamma
\alpha\frac{\xi_{-}(t\mathrm{e}^{k|z|})}{\xi _{-}(t)}\,.
\]
Thus depending on $\gamma \alpha $ either left-handed ($\gamma\alpha =1$)
or right-handed ($\gamma \alpha =-1$) modes are localized on the brane
(again in the sense that the corresponding spinors vanish on the brane).
In particular, for the lowest modes ($n=0$) $\gamma =-1$ the localized
spinor is left-handed if helicity is $\alpha =-1$ and right-handed in the
opposite case.  Making an insight into \eqref{Eq/Pg25/2:fluctpdf0611} we
see that even for the constant mass localization of fermionic
perturbations is possible. The reason for that is based on the special
geometry we work in and on negative curvature of the
space~\cite{Witten:1983ux,Bailin:1987jd}.

\paragraph{Propagation velocity and relation to scalars.}

To conclude this section let us note that from eqs. (\ref{eq:Pg20/1A}) ---
(\ref{eq:Pg20/2}) one sees that  group velocity $\partial E/\partial
p$ in all regimes is greater than unity. Moreover, it follows from eqs.
(\ref{eq:Pg20/1B}), (\ref{eq:Pg20/1C}) and (\ref{eq:Pg20/1}), the group
velocity tends to infinity as $p\to 0$. Thus in order to have equal
velocities at low momenta for the scalar ($c_s=\sqrt{3}$) (\ref{zerodisp})
and the fermionic zero mode one needs to involve some sort of fine tuning.
Namely, one assumes that $m>k/2$ (only in this case the group velocity is
finite), uses (\ref{eq:Pg20/1A}), and solves the  equation
\[
\sqrt{3}=\frac{2m}{2m-k}\,.
\]
The solution is $\frac{m}{k}\simeq 1.18> 1/2$ which confirms that our
assumption is valid.

\section{Conclusions}\label{sec:Conclusions}

In this work we investigated perturbation spectra of the scalar and the
spin-$\half$ fermion fields in the extra dimensional setups with
broken Lorentz invariance in the bulk solutions found in
\cite{Koroteev:2007yp}. In the beginning we qualitatively discussed the
relation between asymptotic behavior of the bulk metric tensor at infinity
in an extra dimension and localization of higher (non-zero) modes by
analyzing geodesic equation in the given metric. Then for the massless
scalar field in the bulk we performed the spectrum analysis by reducing
the corresponding mode equation to the Schr\"{o}dinger equation with a
potential dependent on the metric and particle momentum. Existence of a
zero mode is given rise by the presence of a delta-well in a potential. We have
constructed the parameter space of the bulk metrics used in
\cite{Koroteev:2007yp} with explicitly broken Lorentz invariance in the
bulk and discussed how spectra of excitations behave in different domains
of this space. It has been  shown that if the $g_{00}$ metric
coefficient increases faster then  $g_{ij}$ as the fifth coordinate tends to
infinity, then higher modes of field fluctuations are
\textit{not localized} and the spectrum is continuous; if the opposite case
is realized then a spectrum is discrete and all modes are
\textit{localized} on the brane. Localization properties may, however,
depend on three-dimensional momentum. In the end of \secref{sec:Scalars}
we have elaborated on the model with $g_{00}=1$ and
$g_{ij}=\mathrm{e}^{-2k|z|}\delta _{ij}$ as contrasted to the model
\cite{Dubovsky:2001fj}, in which $g_{00}$ is warped while $g_{ij}$ is not.

The static potential for the massless scalar field was derived. Unlike the RS2
case our potential acquires different power correction at small distances
which could make the two models experimentally distinguishable.

The dispersion relation for the zero mode was obtained and appeared to
break Lorentz invariance explicitly. The speed of scalar massless
particles propagating along the brane (which is governed by the zero mode
dispersion relation) was found to be larger then the speed of tightly
bound massless particle on the brane. Recall that in
\cite{Dubovsky:2001fj} the speed was smaller and one can expect,  that for
intermediate values of Lorentz violating parameters a speed of scalar
particles will be bounded by our result from above and by the result
\cite{Dubovsky:2001fj} from below. However, for completeness, other
domains of the parameter space \figref{fig:xizeta} have to be analyzed.

In \secref{sec:Fermions} spinor fluctuations were investigated on the
example of the model A. We have explicitly found the solution of the Dirac
equation with the mass term $m_{\psi }=m\cdot \mbox{sign}(z)$ and derived
the corresponding dispersion relations. We found fermionic spectra for
different values of the ratio $m/k$. Generically these spectra appeared to
be almost the same as in the scalar case except for the zero mode
dispersion relation. For $m>\half k$ the energy $E$ is proportional to the
momentum $p$ with the coefficient dependent on this ratio; for $m=\half k$
we had $E\sim p \log p$; and $m /k$-dependent power law behavior for
$m<\half k$. Moreover, it turns out that all modes including the zero mode
are polarized but nonchiral\footnote{The modes correspond to a mixture of
left-handed and right-handed spinors with certain coefficients.} unlike the
usual Lorentz invariant case where an appeared zero mode is always
chiral. Thus the energy spectrum is non-degenerate. From the
four-dimensional point of view the nonchiral character of the zero mode
means that it is not described by the Weyl equation --- there is a mixing
between left-handed and right-handed spinors (mass-like term) which
 depends on three-dimensional momentum and disappears in the limit
of small momenta. Apart from that, chirality is precisely restored on the
brane since either left or right-handed components of spinor wave functions
vanish on the brane (the same is true for the higher modes as well).

We have also investigated the case $m_{\psi}=m$. It has an
illustrative character due to the broken $\Integers_{2}$ bulk symmetry.
One can take the limit $m\to 0$ and $\Integers_{2}$ symmetry is
restored. Unlike all the other known braneworld models, a zero mode
\textit{does exist} even for $m=0$ which can be considered as a
consequence of our special geometry and negative curvature of the
space\cite{Witten:1983ux,Bailin:1987jd}. Namely, all excitation modes can
be treated as zero modes since for all modes energy vanishes as
$p\to 0$, unlike the scalar massless case where there is a gap $3k/2$
between zero and higher modes. However, the modes are not polarized and
the energy levels are doubly degenerate. Moreover, all modes are
\textit{nonchiral} as in the $m_{\psi }=m\cdot \mbox{sign}(z)$ case.

Both in the scalar and fermion cases we found that the spectra of higher
modes have very peculiar properties --- at small momenta energies of all
modes tend to one and the same value independent of the mode number. From
the four-dimensional point of view it means that all excitations have the
same mass. However, on the other hand a particle with precisely zero
momentum becomes delocalized and may freely escape from the brane. It is
worth stressing that for a particle somehow localized in the
three-dimensional space (wave packet) the momentum cannot be exactly equal
to zero due to the uncertainty principle.

To conclude, the considered model possesses a lot of unusual and
remarkable properties which are interesting from theoretical as well
as from phenomenological points of view.

\paragraph{Note added.} Recently we found an interesting paper \cite{Kachru:2008yh} where the authors deal with metric deformations which break Lorentz invariance similar to those considered in our work. In \cite{Gordeli:2009vh} an explicit comparison of the two solutions is made. The work \cite{Kachru:2008yh} represents a prospective programme of gauge/gravity correspondence for theories with broken Lorentz invariance.

\section*{Acknowledgements}

The authors are grateful to S. Dubovsky, D. Gorbunov, J. Kinney, D. Levkov, S.
Si\-bi\-rya\-kov, F. Spill, A. Zayakin for fruitful discussions and
especially to V. Rubakov and A. Vainshtein for reviewing the final version
of the manuscript. M.L. would like to thank Service de Physique
Th\'{e}orique, Universit\'{e} Libre de Bruxelles where part of this work
was done under support in part by the Belspo:IAP-VI/11 and IISN grants,
for kind hospitality. P.K. thanks Max-Planck-Institut f\"ur
Gravitationsphysik (Albert-Einstein-Institut) where part of his
work was done for support and hospitality.

This work is partly supported by the grant of the President of Russian
Federation (NS-1616.2008.2), the RFBR grant 08-02-00473 and by the grant
of Dynasty Foundation awarded by the Scientific Board of ICFPM.

\appendix
\section{Appendix}
\label{sec:Derivation}

In this Appendix we give detailed derivations of the dispersion relations
(\ref{eq:Pg20/1A}) -- (\ref{eq:Pg20/2}) for the fermionic spectrum. For
simplicity we introduce the following dimensionless variables $t=p/k$,
$\varepsilon =E/k$, and $\mu =m/k$. We start from the
equation~(\ref{Eq/Pg20/1A:fluctpdf0611}).

\paragraph{Small momenta.}

Let us firstly discuss the regime of small momenta $t\ll 1$. One can
expand the l.h.s. of (\ref{Eq/Pg20/1A:fluctpdf0611}) at small $t$ and
obtain the following equation
\begin{equation}\label{eq:MacpmExpansion}
\left[1+\frac{t}{2\nu -1} +\frac{\Gamma (1/2-\nu )}{\Gamma (1/2+\nu )}
\left(\frac{t}{2}   \right)^{2\nu } \right]=\gamma \frac{\mu -\alpha
\varepsilon }{\nu }\left[1-\frac{t}{2\nu -1}
-\frac{\Gamma (1/2-\nu )}{\Gamma (1/2+\nu )} \left(\frac{t}{2}
\right)^{2\nu } \right]\,.
\end{equation}
Here we kept two terms in the next to the
leading order since depending on $\nu$ these terms may compete
with each other; the corrections to the equation are of the order of
$t^{2}$ and $t^{2\nu +1}$. If $\Re\nu \neq 0$ (which automatically
implies $\mu \neq 0$) then the leading order is represented by the first
terms in the square brackets in \eqref{eq:MacpmExpansion}. Thus in order to
satisfy the equation one is forced to conclude that $\varepsilon \to 0$ as
$t\to 0$ and $\gamma =1$, which means that we deal with the zero mode.
Aside from that, it turns out that $\alpha $ is negative.
This fact follows straightforwardly from (\ref{Eq/Pg20/1A:fluctpdf0611}).
Indeed, the operator (\ref{Eq/Pg18/2:fluctpdf0611}) with the boundary
conditions under consideration is hermitian, so, if $\Re\nu  \neq 0$ then
$\Im\nu =0$. It is well known that at real values of order the
modified Bessel function $\Mac_{\nu }(t)$ is positively defined and at
fixed argument is an increasing function of $\nu $. Thus, $\xi _{-}(t)>\xi
_{+}(t)>0$ what means that $\alpha =-1$ and $\gamma =1$.

In order to find the behavior of the zero mode at $t\to 0$ one needs to take into
account subleading terms in \eqref{eq:MacpmExpansion}. Expanding
(\ref{eq:MacpmExpansion}) at small $\epsilon $ and $t$ one obtains
\begin{equation}
\label{eq:EnergyZeroGeneral}
\epsilon\simeq
\frac{2\mu}{2\mu-1}t+2\mu\frac{\Gamma(\half-\mu)}{\Gamma(\half+\mu)}\left(\frac{t}{2}\right)^{2\mu}\,.
\end{equation}
which reconstructs all dispersion relations for zero mode
(\ref{eq:Pg20/1A}) -- (\ref{eq:Pg20/1C}) at small momenta. Indeed, if $\mu
>1/2$ then the first term in (\ref{eq:EnergyZeroGeneral}) is dominant, the
second term is dominant in the opposite case ($\mu <1/2$). In the case
$\mu \to 1/2$ both terms tend to infinity becoming comparable. Expanding the
r.h.s. of (\ref{eq:EnergyZeroGeneral}) at $\mu =1/2$ one easily
verifies that divergent parts are cancelled and obtains \eqref{eq:Pg20/1B}.

Recall that in the above analysis we assumed $\Re\nu\neq 0$. Let us now
consider the case when $\nu$ is pure imaginary. In this
case the leading terms in the brackets in (\ref{eq:MacpmExpansion}) are the
first and third one. Then after some calculations one obtains from
(\ref{eq:MacpmExpansion}) the following equation (the same equation
follows directly from (\ref{eq:ChiMatching}))
\begin{equation}
\left(\frac{t}{2}   \right)^{-2\nu }\Gamma \left(\frac{1}{2} +\nu
\right)^{2}(\gamma \mu -\nu )- \left(\frac{t}{2}   \right)^{2\nu }\Gamma
\left(\frac{1}{2} -\nu \right)^{2}(\gamma \mu +\nu )=0\,.
\label{Eq/Pg31/1:fluctpdf0611}
\end{equation}
which is equivalent to
\begin{equation}
\label{Eq/Pg31/2:fluctpdf0611}
-2\nu
\log\frac{t}{2}+2i\arctan\frac{1}{i}\frac{\Gamma (1/2+\nu )-\Gamma
(1/2-\nu ) }{\Gamma (1/2+\nu )+\Gamma (1/2-\nu )} +i\arctan\left[\gamma
\frac{i\nu }{\mu } \right]=\pi n i\,,
\end{equation}
where n is an integer. Since the last two terms in the l.h.s. of this
equation are finite at all values of imaginary $\nu$, the only way to
satisfy the equation at fixed $n$ is to assume that $\nu$ thends to zero as
$t\to0$. Note, however, that the expression (\ref{Eq/Pg31/2:fluctpdf0611})
does not depend on $\alpha$, i.e. it holds for both helicities. In order to
clarify the dependence on $\alpha$ one should find $\nu $ directly from
(\ref{eq:MacpmExpansion}) (the advantage of (\ref{Eq/Pg31/1:fluctpdf0611})
is that this equation allows one to establish quite easily the fact that
$\nu\to 0$ as $t\to 0$, while it is not so obvious from
(\ref{eq:MacpmExpansion})).

Performing small $\nu$ expansion in (\ref{eq:MacpmExpansion}) one obtains
\begin{equation}
\label{Eq/Pg34/1:fluctpdf0611}
\tanh\left(\nu \Psi
\left(\half\right)-\nu\log\frac{t}{2}\right)
=\left(\frac{2\gamma\mu}{\nu}\right)^{\alpha}\,.
\end{equation}
This equation tells us that if $\alpha =1$ then the argument of $\tanh$ should
go to $i\pi (l+\half),\,l\in\Integers$, meanwhile if $\alpha=-1$ the
argument tends to $i \pi l $. Then the solution of
(\ref{Eq/Pg34/1:fluctpdf0611}) can be written as follows
\begin{equation}
\label{eq:Pg28/1}
\nu_{n}(t)=\frac{i\pi n}{2\Psi(\half)-
\displaystyle\frac{\gamma }{\mu}-2\log\frac{t}{2}}\,,
\end{equation}
where $n\in\mathbb{N}$ (recall that $\nu \neq 0$ and we assume that
$\Im(\nu )>0$),
and mode helicities depend on the level number as $\alpha=(-1)^{n+1}$. The
above equation in turn gives us the dispersion relation for the heavy
modes (\ref{eq:Pg20/1}). Note that (\ref{eq:Pg28/1}) is obtained in the
assumption  $|\nu|\ll 1$. Thus for $n$ large enough the momentum $t$
should be small. However, it does not affect the conclusion that the
spectrum is discrete and $E_{n} \to m$ as $t\to 0$.

Let us now consider the case $m_{\psi }=m$, i.e. we start from
(\ref{Eq/Pg21/1:fluctpdf0611}). Comparing (\ref{Eq/Pg21/1:fluctpdf0611})
with (\ref{Eq/Pg20/1A:fluctpdf0611}) one sees that
(\ref{eq:MacpmExpansion}) still holds if one replaces the factor
$(\mu-\alpha\varepsilon)/\nu $ in the r.h.s. by the factor of $i$ which is
equivalent to taking $\mu=0$ and $\alpha =1$ in this (and only in this)
factor. The equation (\ref{Eq/Pg31/1:fluctpdf0611}) also holds if one puts
$\gamma \mu =0$ in the parentheses. So the conclusion that $\nu $ goes to
zero as $t\to 0$ remains intact, i.e. all modes have the same mass
$\mu $, in particular in the massless case all modes become gapless. In the
regime $\nu \to 0$ one can obtain from (\ref{eq:MacpmExpansion}) the
following equation (an analog of (\ref{Eq/Pg34/1:fluctpdf0611}))
\[
\tanh\left(\nu \Psi \left(\frac{1}{2}\right)-\nu
\log\frac{t}{2}\right)=-i\gamma\,.
\]
Then
\begin{equation}
\label{Eqn/Pg34/1:fluctpdf0611}
\nu _{n}(t)=\frac{i\pi
(2n+1)}{4\Psi(\half)-4\log\frac{t}{2}}\,.
\end{equation}
Note that $\gamma$ depends on the mode number $n$ as $\gamma=(-1)^{n+1}$.
From (\ref{Eqn/Pg34/1:fluctpdf0611}) the dispersion relation
(\ref{Eqn/Pg24/1:fluctpdf0611}) immediately follows.

\paragraph{High momenta.}

Let us now consider the high momenta regime $t\gg \mu ,\epsilon$. In this
case it is easy to start from (\ref{Eq/Pg18/2:fluctpdf0611}) and neglect
the last term compared with the term proportional to $p^{2}$. Then the
solutions to the equations of motion  reads
\[
\xi_{\pm}=\Mac_{\nu }(t),\ \ \ \ \nu =\sqrt{\mu ^{2}-\varepsilon
^{2}}\,,
\]
and the equation yielding the dispersion relation (an analog of
(\ref{eq:ChiMatching})) is the following
\begin{equation}
\label{eq:Pg28/3}
t\,\Mac'_\nu(t) =0\,,
\end{equation}
where we neglected the term $\mu \Mac_{\nu }$ which appears in the case
$m_{\psi }=m\cdot\mbox{sign}(z)$. Thus  one can consider both cases
($m_{\psi }=m\cdot\mbox{sign}(z)$ and $m_{\psi }=m$) on an equal footing.

There are three different possibilities in the regime $t\to \infty $: 1)
$|\nu|\ll t$, 2) $|\nu|>t$, 3) $|\nu|\to \infty$ but $|\nu|<t$. In the
first case one can use the asymptotic behavior of the Bessel function with
large argument and fixed order
\begin{equation}
\label{eq:Pg28/4}
\Mac_{\nu}(t)=\sqrt{\frac{\pi}{2t}}\,\mathrm{e}^{-t}
\left(1+\mathcal{O}\left(t^{-1}\right)\right)\,.
\end{equation}
Due to the fact that the leading order in (\ref{eq:Pg28/4}) does not
depend on $\nu$ there is no solution of (\ref{eq:Pg28/3}) in this case. In
the second and in the third cases $\varepsilon \to \infty $ as $t\to
\infty $, so $\nu$ is pure imaginary $\nu =i|\nu|$. In the second case one
uses the  asymptotic expansion for the Bessel function
\cite{Bateman:1953}
\[
\Mac_{i|\nu|}(t)=\frac{\sqrt{2}}{\sqrt[4]{|\nu|^{2}-t^{2}}}
\mathrm{e}^{\frac{|\nu |\pi }{2}}\left[\Gamma\left(\half \right)
\sin\left(|\nu
|\mbox{Arccosh}\frac{|\nu|}{t}-\sqrt{|\nu|^{2}-t^{2}}+\frac{\pi}{4}\right)
+\mathcal{O}\left(t^{-1}\right)\right]\,.
\]
Substituting this expression into (\ref{eq:Pg28/3}) one obtains the
 equation
\[
\tan\left(|\nu|\mbox{Arccosh}\frac{|\nu
|}{t}-\sqrt{|\nu|^{2}-t^{2}+\frac{\pi }{4}} \right)=
2\frac{\left(|\nu|^2-t^2\right)^{\sfrac{3}{2}}}{t^{2}}\,.
\]
The solution of this equation reads
\begin{equation}
\label{eq:Pg28/5A}
|\nu|=t+\mathcal{O}(t^{1/3})\,,
\end{equation}
where the correction is positive, so the condition $|\nu|>t$ is satisfied.
From this equation one obtains (\ref{eq:Pg20/2}).

In the last case $|\nu|<t$ one uses the  asymptotic
expansion~\cite{Bateman:1953}
\[
\Mac_{i|\nu |}=\frac{\Gamma\left(\half\right)+\mathcal{O}\left(t^{-1}
\right)}{\sqrt[4]{4(t^{2}-|\nu|^{2})}}\exp\left( -\sqrt{t^{2}-|\nu
|^{2}}-|\nu |\arcsin\frac{|\nu|}{t} \right) \,.
\]
Substituting this expansion in (\ref{eq:Pg28/3}) one again ends up with
the solution (\ref{eq:Pg28/5A}) where the correction is positive as well.
It contradicts however the assumption $|\nu|<t$. Thus there is no solution
of (\ref{eq:Pg28/3}) in this case.


\bibliography{fluct}
\bibliographystyle{aei}

\end{document}